\documentclass[footinbib,twocolumn,amstex,mathfonts,prl,10pt]{revtex4-1}
\usepackage{graphicx,bm,verbatim,soul,caption,bbm}
\DeclareCaptionJustification{justified}{\leftskip=0pt \rightskip=0pt \parfillskip=0pt plus 1fil}
\usepackage[usenames, dvipsnames]{color}
\usepackage[pdftex,colorlinks=red,bookmarks=false]{hyperref}
\usepackage{amsmath,amssymb,amsthm,amsfonts,float}
\usepackage{enumitem}
\usepackage[caption=true]{subfig}
\begin{document}
\def\qv{\vec{q}}

\def\blue{\textcolor{black}}
\newcommand{\norm}[1]{\left\lVert#1\right\rVert}
\newcommand{\ad}[1]{\text{ad}_{S_{#1}(t)}}

\newcommand{\mi}{\mathrm{i}}
\newcommand{\me}{\mathrm{e}}

\title{Tidal surface states as fingerprints of non-Hermitian nodal knot metals}

\author{Xiao Zhang$^{1}$}
\author{Guangjie Li$^{1,2}$}
\author{Yuhan Liu$^{1,3}$}
\author{Tommy Tai$^{4}$}
\author{Ronny Thomale$^5$}
\email{rthomale@physik.uni-wuerzburg.de}
\author{Ching Hua Lee$^{6}$}
\email{phylch@nus.edu.sg}
\affiliation{$^1$School of Physics, Sun Yat-sen University, Guangzhou 510275, China.}
\affiliation{$^2$Department of Physics and Astronomy, Purdue University, IN 47907-2036, USA.}
\affiliation{$^3$Department of Physics, the University of Chicago, Chicago, Illinois 60637, USA.}
\affiliation{$^4$Cavendish Laboratory, University of Cambridge, JJ Thomson Ave, Cambridge CB3 0HE, United Kingdom}
\affiliation{$^5$Institut f\"{u}r Theoretische Physik und Astrophysik,
  Universit\"{a}t W\"{u}rzburg, Am Hubland Campus S\"{u}d,
  W\"{u}rzburg 97074, Germany}
\affiliation{$^6$Department of Physics, National University of Singapore, Singapore 117542.}
\date{\today}

\begin{abstract}
\section*{Abstract}
Non-Hermitian nodal knot metals (NKMs) contains intricate complex-valued energy bands gives rise to knotted exceptional loops and new topological surface states. We introduce a formalism that connects the algebraic, geometric, and topological aspects of these surface states with their parent knots, and provide an optimized constructive ansatz for tight-binding models for non-Hermitian NKMs of arbitrary knot complexity and minimal hybridization range. Specifically, various representative non-Hermitian torus knots Hamiltonians are constructed in real-space, and their nodal topologies studied via winding numbers that avoid the explicit construction of generalized Brillouin zones. In particular, we identify the surface state boundaries as ``tidal'' intersections of the complex band structure in a marine landscape analogy. Beyond topological quantities based on Berry phases, we further find these tidal surface states to be intimately connected to the band vorticity and the layer structure of their dual Seifert surface, and as such provide a fingerprint for non-Hermitian NKMs.
\end{abstract}

\maketitle

\section{Introduction}
Since the early days of a quantum theory of metals, surface states have been appreciated as fundamental phenomena arising from the surface terminations experienced by electronic waves~\cite{tamm}. With the advent of topological nodal semimetals~\cite{PhysRevB.84.235126,RevModPhys.88.035005,RevModPhys.90.015001}, not only geometry but also bulk topology has emerged as a source for metallic surface states, linking bulk topology and surface state profile~\cite{PhysRevB.94.045113}. Non-Hermiticity has been appearing as yet another level of differentiation and complexity that intertwines topology and metallicity for not just quantum electrons, but also classical analogs.  
In those contexts, metallicity refers not to a Fermi surface intersection,
but essentially embodies the antithesis of a band insulator, i.e. the absence of spectral gaps~\cite{PhysRevX.9.041015} Besides parity-time (PT)-symmetric systems with real eigenspectrum due to balanced gain and loss~\cite{feng2017non}, non-Hermiticity, in combination with topology and surface terminations, has been recently shown to unfold a rich scope of experimentally robust phenomena far beyond mere dissipation~\cite{zhen2015spawning,st2017lasing,weimann2017topologically,leykam2017edge,gao2018chiral,zhou2018observation,PhysRevB.98.205417,PhysRevLett.123.066405,PhysRevB.102.035101,lee2020ultrafast,  PhysRevLett.124.250402,PhysRevB.100.155117,zhao2019non,li2020impurity,borgnia2020non} (see \cite{Ghatak_2019,lu2014topological,bergholtz2019exceptional,ashida2020non,de2017dynamics} for excellent reviews).

One active arena particularly fueled by analyzing non-Hermiticity is the quest for exotic phases in nodal knot metals~\cite{bi2017nodal,ezawa2017topological,PhysRevB.99.161115,wang2017scheme,unal2020topological,yi2019observation} (NKMs), whose intricate knotted topology~\cite{dowker1983classification,dennis2010isolated} 
in 3D transcends traditional $\mathbb{Z}$ and $\mathbb{Z}_2$ classifications~\cite{ryu2010topological}. Non-Hermiticity lifts the requirement of sublattice symmetry, leading to more robust NKMs which, as we will show, can also be practically realized due to rather short-ranged couplings. 

Yet, despite their principal appeal, key aspects of non-Hermitian NKMs remain poorly understood. 

Specifically, no systematic understanding of the shape, location, and topology of non-Hermitian NKM surface state regions currently exists beyond isolated numerical results~\cite{PhysRevB.99.081102,PhysRevB.99.075130,PhysRevB.99.161115}. This conceptual gap has endured until today, because non-Hermiticity modifies the topological bulk-boundary correspondence in subtle complex-analytic ways, which so far have not been studied beyond 1D, especially for models that possess complicated sets of hoppings across various distances~\cite{lee2016anomalous,PhysRevB.99.201103,yao2018edge,xiong2018does, PhysRevLett.124.086801,PhysRevB.102.081115,longhi2019probing,Li2020,zhang2020non}.

In this work, we devise a comprehensive formalism that relates surface states of non-Hermitian NKMs to their Seifert surface (knot) topology, complex geometry, vorticity, and other bulk properties. Each of these properties have separately aroused much interest: Knot topology concerns the innumerable distinct configurations of knots, so intricate that they cannot be unambiguously classified by any single topological invariant; the complex analytic structure of bandstructures have led to various non-Hermitian symmetry classifications which are also augmented by the non-Hermitian skin effect; and half-integer vorticity underscores the double-valuedness around exceptional points. The study of models with complicated hoppings across varying distances often involve the evaluation of generalized Brillouin zones (GBZ)~\cite{PhysRevB.102.085151, yang2020non}, but in studying their nodal topologies, we shall adopt a winding number counting approach that avoids the need for such tedious or non-analytic computations.

 Unlike previous works~\cite{PhysRevB.99.081102,carlstrom2018exceptional,PhysRevB.99.075130}, we shall be primarily concerned with non-Hermitian NKMs not perturbatively connected to known Hermitian analogs. Despite their sophistication, these NKMs exhibit short-ranged tight-binding representations potentially realizable in disordered semimetals and non-reciprocal electrical or photonic circuits~\cite{feng2011nonreciprocal,tzuang2014non,PhysRevLett.122.247702,luo2018nodal,PhysRevB.99.041116, PhysRevResearch.2.023265,lee2020imaging,Helbig2020,Imhof2018,Lee2018}. In particular, we illustrate how the topological tidal surface states can be mapped out as topolectrical resonances in non-Hermitian circuit realizations, based on recent experimental demonstrations involving analogous 1D circuit arrays~\cite{Helbig2020}.

Core to our formalism is the interpretation of non-Hermitian pumping as a ``tidal'' movement in a marine landscape analogy of the complex band structure. 
In this picture, familiar Hermitian NKM topological ``drumhead'' regions~\cite{weng2015topological} become special cases of generic ``tidal'' islands that determine the surface state regions in both Hermitian and non-Hermitian cases. In particular, we present a direct link between the 2D surface tidal states and the Seifert surface bounding a 3D dual NKM, which encapsulates full nodal topological information. The tidal island topology, which here refers to the connectedness of its regions, corresponds directly to the layer structure of its dual Seifert surface. 
Interestingly, the interplay between surface-projected nodal loops (NLs) and the tidal regions also constrain the vorticity and hence the spectral cobordism\cite{luck2002basic} along particular Brillouin Zone (BZ) paths. Evidently, all these phenomena do not exist in 1D or 2D non-Hermitian systems, and thus illustrate deep connections between topological protection and non-Hermitian pumping that manifest only in higher dimensions. Our results thus provide a fingerprint for non-Hermitian NKMs.

\section{Results} 

\subsection{Models for non-Hermitian NKMs } We first introduce an ansatz for NKMs representing the important class of $(p,q)$-torus knots. Here, we specialize to knots that can be represented as a loop on the surface of a torus. Specifically, a $(p,q)$-torus knot is one that winds $p$ times around the symmetry axis while also winding $q$ times around the internal circle direction. These knots are isomorphic to closed braids with $p$ strands each twisting $q$ times around a torus, with the number of linked loops being the greatest common divisor (GCD) of $p$ and $q$, i.e. $\text{GCD}(p,q)$ linked loops~\cite{murasugi2007knot,bode2017knotted,lee2020imaging,li2019emergence}, and encompasses many common knots like the Hopf-link and the Trefoil knot. Despite their seeming geometric and topological complexity, we shall see that the enlarged set of non-Hermitian coefficients allows for rather local implementations of these nodal structures.

A minimal nodal Hamiltonian consists of two bands:
\begin{equation}
H(\bold k)=h_x(\bold k)\sigma_x+h_y(\bold k)\sigma_y+h_z(\bold k)\sigma_z=\bold h(\bold k)\cdot \mathbf{ \sigma},
\label{Hxyz}
\end{equation}
where $\bold k\in \mathbb{T}^3$. Nodes (gap closures) occur when $h_x^2+h_y^2+h_z^2=|\text{Re}\,\bold h|^2-|\text{Im}\,\bold h|^2+2i\,\text{Re}\,\bold h\cdot \text{Im}\,\bold h=0,$ 
which is equivalent to the two conditions $|\text{Re}\,\bold h|=|\text{Im}\,\bold h|$ and $\text{Re}\,\bold h\cdot \,\text{Im}\,\bold h=0$. Thus, nodal loops generically exist in 3D as as long as $H(k)$ is non-Hermitian ($\text{Im}\,\bold h\neq 0$). Inspired by constructions of Hermitian NKMs, we employ the ansatz~\cite{Methods}
\thickmuskip=1mu
\begin{equation}
\bold h=\left\{\begin{array}{lll}
(2\mu^i,2w^j,0),& &p=2i,q=2j\\
(2\mu^i,w^j+w^{j+1},\gamma),& &p=2i,q=2j+1\\
(\mu^i+\mu^{i+1},w^j+w^{j+1},\gamma),& &p=2i+1,q=2j+1,
\end{array}\right.,
\label{hpq}
\end{equation}
where $\gamma\approx\,i$ is empirically tuned to ensure the desired crossings, and 

\begin{eqnarray}
\mu(\bold k)&=&\sin k_3 + i(\cos k_1 +\cos k_2+\cos k_3-m),\qquad\notag\\
w(\bold k)&=&\sin k_1 + i\sin k_2, \qquad \quad \,\,\, 1.5<m<2.5.\qquad,
\label{muw}
\end{eqnarray}
the range of $m$ also empirically constrained to prevent the appearance of extraneous solutions not belonging to any knot. Due to the freedom in having complex components in $\bold h(\bold k)$, Eq.~\ref{hpq} contains hybridizations across at most $\text{max}\left(\lceil\frac{p}{2}\rceil,\lceil\frac{q}{2}\rceil\right)$ unit cells, approximately half of the $\text{max}(p,q)$ unit cell range of their Hermitian counterparts (Fig.~\ref{fig:braid}(a-d))~\cite{ezawa2017topological,bi2017nodal,bode2019constructing,tai2020anisotropic}. 

\subsection{Topological (Tidal) surface states} 
\begin{figure}
\centering
\includegraphics[width= \linewidth]{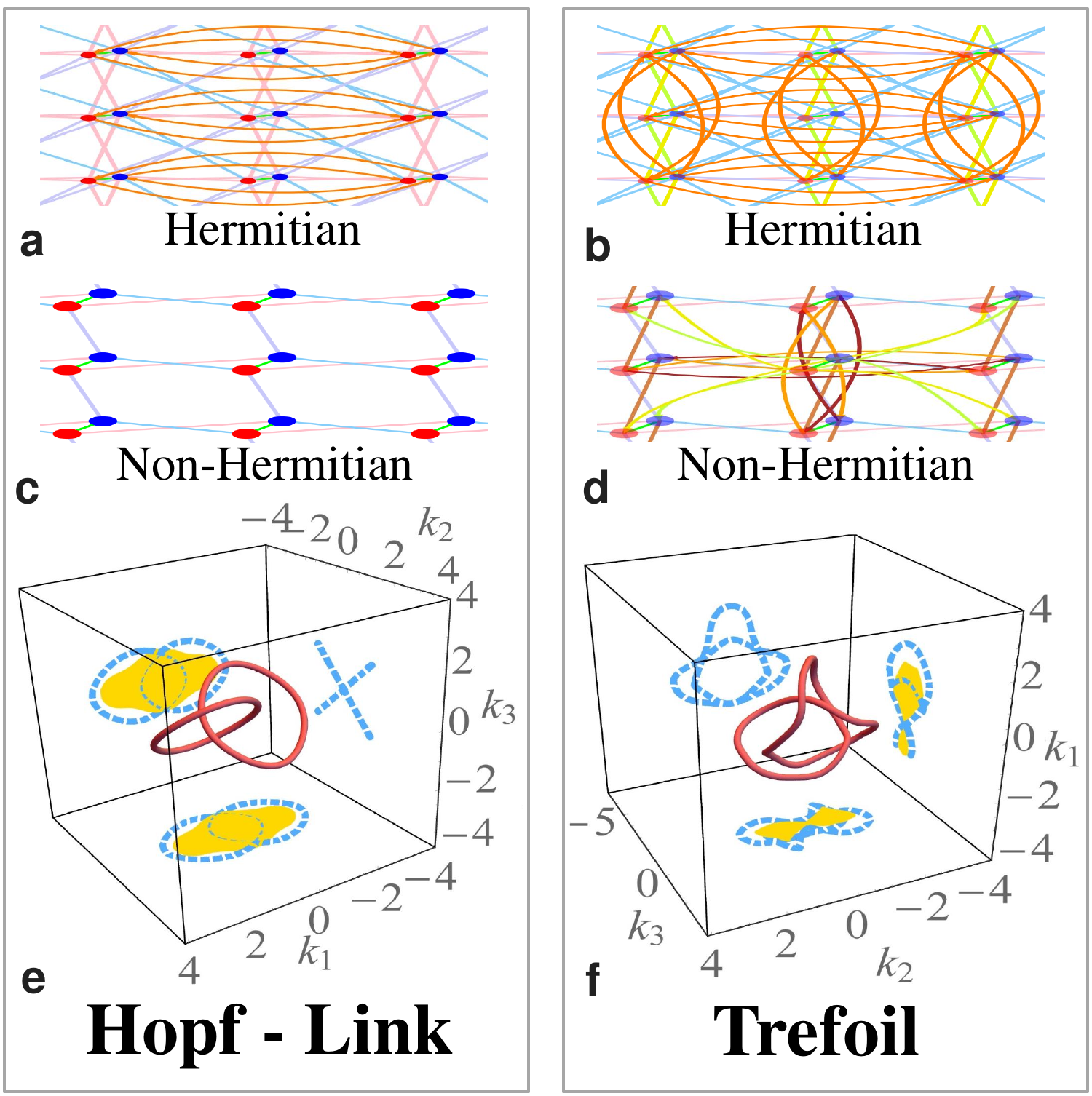}
\caption{Comparison of the Hopf link and Trefoil knot  nodal knot metals (NKMs).}
\caption*{We compare the two simplest nodal knot metals (NKMs) - the Hopf link (a,c,e) and the Trefoil knot (b,d,f), constructed from Eqs.~\ref{Hxyz},\ref{hpq},\ref{muw}, with the parameter $m$ set to 2. The former is a (2,2)-torus knot, while the latter is a (2,3)-torus knot. The tight-binding implementation~\cite{Methods} shows the enhanced short-rangedness of the non-Hermitian model (c-d), as compared to the Hermitian model (a-b). The surface state projections (e-f) highlight the difference between the usual ``drumhead'' regions demarcated by the surface projections (blue dashed curves) of the bulk nodal lines (red) for the Hermitian case vs. the non-Hermitian topological ``tidal'' states (yellow). }
\label{fig:braid}
\end{figure}

Unlike Hermitian nodal systems, our non-Hermitian NKMs exhibit topological surface state regions not bounded by surface projections of the bulk NLs (``drumhead" boundaries). Rather, they are shaped like ``tidal'' regions (Fig.~\ref{fig:braid}(e-f)), a nomenclature which will be elucidated shortly. This is unlike usual Hermitian nodal structures, where the drumhead surface states are so-named because they are bounded by surface projections of the bulk NLs. The underlying reason is that as we move from periodic to open boundary conditions (PBCs to OBCs), the effective perpendicular couplings are generically asymmetric, causing macroscopically many eigenstates, including former bulk states, to accumulate at the boundaries and form ``skin" states~\cite{alvarez2018non,PhysRevB.99.201103,yao2018edge}.  
As such, it is the gap closures of the skin states, not bulk states, that determine topological phase boundaries. While the skin effect per se has been well-studied in 1D, the beautiful relations of its boundary states with vorticity, complex band structure and Seifert surfaces in a higher-dimensional nodal setting are what we intend to uncover in this work.

\noindent

Consider a surface normal $\hat n$, and define normal and parallel momentum components $k_\perp$ and $\bold k_\parallel=\bold k-k_\perp \hat n$. The existence of topological surface states depend on the off-diagonal components of the NKM Hamiltonian 
\begin{comment}
\begin{equation}
H(\bold k)=\left(\begin{matrix}
h_z & h_x-i k_y \\
h_x+ i h_y & -h_z
\end{matrix}\right)
=\left(\begin{matrix}
h_z() & a(z) \\
b(z) & -h_z(z)
\end{matrix}\right)
\end{equation}
\end{comment}
specified by Eqs.~\ref{Hxyz} to \ref{muw}, which are are most conveniently parametrized by (with $z=e^{i k_\perp}$):
\begin{align}
a(z;\bold k_\parallel)&=h_x(\bold k)-ih_y(\bold k)=\tilde az^{r_a-p_a}\prod_i^{p_a} (z-a_i) \notag\\
b(z;\bold k_\parallel)&=h_x(\bold k)+ih_y(\bold k)=\tilde bz^{r_b-p_b}\prod_i^{p_b} (z-b_i) ,
\label{Ham}
\end{align}
\begin{figure}
\centering
\includegraphics[width= \linewidth]{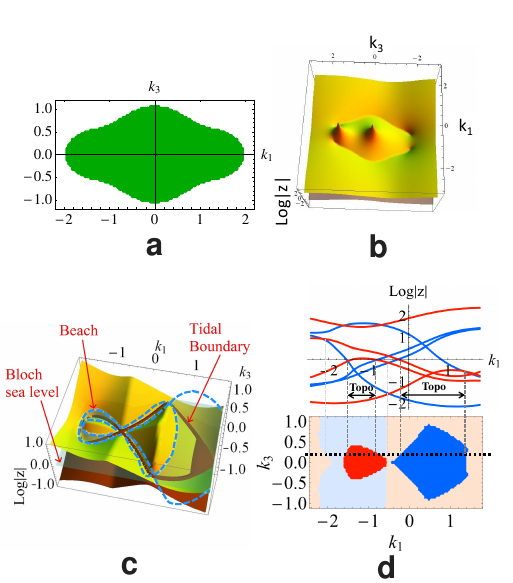}
\caption{Topological (tidal) surface states and imaginary gap band structure ($\log|z|$ where $z=e^{ik_2}$) of the non-Hermitian nodal Hopf-link (a-b) and Trefoil knot (c-d).}
\caption*{(a) Topological (tidal) surface states (normal to $\hat e_2$) of the non-Hermitian nodal Hopf-link, with analytic expression given by Eq.~\ref{tptm}.  
(b-c) Plot of the imaginary gap ($\log|z|$ where $z=e^{ik_2}$) band structure across the $\hat e_2$ surface Brillouin Zone (BZ) of the non-Hermitian Hopf link (b) and Trefoil knot (c). The analytically obtained tidal region of (a) are exactly demarcated by the trenches (tidal boundaries) in (b). (c) Topological (tidal region) boundaries are the intersections (trenches) between its 4-th (yellow) and 5-th (brown) bands.
The periodic boundary conditions (PBC) spectrum gap closes along band intersections (dashed blue beaches) with the cyan ``sea level" surface $\log|z|=0$ (which has real $k_2=-i\log z$).
(d) Correspondence between the tidal phase boundaries in $k_1$-$k_3$ space, and the $r_a+r_b=4$ highest $a_i(\bold k_\parallel)$ (blue) and $b_i(\bold k_\parallel)$ (red) $\log|z|$ bands along an illustrative $\bold k_\parallel=(k_1,k_3=0.2)$ line. Case 3-type intersections between the 4-th and 5-th bands mark tidal phase boundaries with blue(red) regions corresponding to the 4-th band being $a_i$($b_i$), shaded (lightly)brightly according to whether the topological criterion Eq.~\ref{winding} is (not)satisfied. 
}
\label{fig:XX2}
\end{figure}

\noindent where $r_a,r_b,p_a,p_b$ are integer exponents and $\tilde a,\tilde b,a_i,b_i$ are functions of $\bold k_\parallel$ determined by the model being studied. Since $\bold k_\parallel$ coordinates are just spectators in taking the OBCs, they can be regarded as parameters indexing a collection of 1D OBC chains along $\hat n$. Most generically, surface topological modes exist at $\bold k_\parallel$ where there exists a contour $|z|=R$ such that the windings~\cite{PhysRevB.99.201103} (where $a'(z) = da/dz$ and similarly for $b'(z)$) are
\begin{equation}
\Gamma_{a}^{\bold k_\parallel,R}=\oint_{|z|=R} \frac{a'(z;\bold k_\parallel)\,dz}{i\,a(z;\bold k_\parallel)},\,\,\, \Gamma_{b}^{\bold k_\parallel,R}=\oint_{|z|=R} \frac{b'(z;\bold k_\parallel)\,dz}{i\,b(z;\bold k_\parallel)}\quad
\label{windings_ab}
\end{equation}
have opposite signs, i.e. in the topological region, 
\begin{equation}
\exists\, R\in(0,\infty) \ \ \ \text{such that}\ \   \Gamma_{a}^{\bold k_\parallel,R}\,\Gamma_{b}^{\bold k_\parallel,R}<0.
\label{winding}
\end{equation}
\thickmuskip=3mu

\noindent This is the criterion to determine topological surface states. As these conditions are not the same as that for Hermitian PH symmetric topological modes, we expect the NKM topological regions to be different from the drumhead regions of Hermitian nodal metals. We shall call them ``tidal regions'', since we will see that they can be intuitively understood via ``tidal effects'' in a marine landscape analogy. As the underlying arguments are rather intricate, we shall first elaborate on the simplest example of the nodal Hopf link, and then show how that motivates such a graphical interpretation.

We first demonstrate how to obtain the topological (tidal) region for the simplest non-Hermitian NKM, the Hopf link. From Eq.~\ref{hpq}, a nodal Hopf link can be defined by $\bold h=(2z,2w,0)$, so that for a surface normal to $\hat n =\hat e_2$, $\bold k_\parallel=(k_1,k_3)$. We have $a(z;\bold k_\parallel)=2i\left(z^{-1}-t_+\right)$ and $b(z;\bold k_\parallel)=2i\left(z-t_-\right)$, which is just the non-reciprocal SSH model~\cite{yao2018edge,PhysRevB.99.201103,PhysRevB.99.081103} with dissimilar intra/inter-unit cell hopping ratios $t_\pm=m\pm\sin k_1-\cos k_1-e^{-ik_3}$. Using Eqs.~\ref{Ham} to~\ref{winding} (with $p_a=p_b=1$ and $r_a=1,r_b=0$), we obtain $\Gamma_{a}^{\bold k_\parallel,R}=-\theta(t_+^{-1}-R)$ and $\Gamma_b^{\bold k_\parallel,R}=\theta(R-t_-)$, such that the topological region (yellow in Fig.~\ref{fig:braid}(e-f)) is given by the set of $(k_1,k_3)$ satisfying~\cite{Methods}
\medmuskip=1mu
\begin{equation}
|t_+t_-|=\prod_\pm\left[(m\pm\sin k_1-\cos k_1-\cos k_3)^2+\sin^2 k_3\right]<1.
\label{tptm}
\end{equation}
This region, as illustrated in Fig.~\ref{fig:braid}e,\ref{fig:XX2}a, is qualitatively different from the usual Hermitian drumhead states, which are ``shadows'' of the Hopf loops $|t_+|=1$ and $|t_-|=1$ (blue dashed outlines in Fig.~\ref{fig:braid}) on the OBC surface. 

While the shape of this topological region can be analytically derived and written down (Eq.~\ref{tptm}) for this simplest Hopf link example, more insight can be obtained from the imaginary gap~\cite{he2001exponential,lee2015free,lee2016band} band structure, which is the solution set of $z=e^{ik_\perp}$ that closes the gap, i.e. values of $z$ (bands) that solves $\text{Det}\,H(z;\bold k_\parallel) =0$ for each $ \bold k_\parallel$. 

The imaginary gap solutions of the Hopf link model is shown in Fig.~\ref{fig:XX2}b. Notice that its surfaces intersect precisely along the boundary of the region with topological surface states (Fig.~\ref{fig:XX2}a)! This is a result of the equivalence between Eqs.~\ref{windings_ab} - \ref{winding} and the surrogate Hamiltonian formalism~\cite{PhysRevB.102.085151}. Indeed, when the imaginary gap solutions intersect, the skin mode solutions also experience gap closure, thereby allowing for topological transitions.

For generic NKMs, however, there exist $p_a+p_b$ imaginary gap solutions, and a more sophisticated graphical treatment is necessary. This is why while biorthogonal arguments from Ref.~\onlinecite{kunst2018biorthogonal} can produce similar results in the Hopf link example, they will be inconclusive in generic nodal systems where $p_a,p_b>1$, such that multiple roots exist for $a(z,\bold k_\parallel)$ and $b(z,\bold k_\parallel)$. For instance, the Trefoil NKM has $8$ imaginary gap solutions, and it is the intersection between the 4th (yellow) and 5th (brown) imaginary gap solutions (bands) (Fig.~{\ref{fig:XX2}c) that demarcates the $\hat e_2$ surface state boundaries (Fig.~\ref{TrefoilBandStructure}b and \ref{fig:XX3}(a,c)). 

In this picture, the (cyan) sea level at $\log|z|=0$
 keeps track of Bloch states with real $k_\perp=-i\log z$, with the intersections (blue dashed ``beaches") of the sea level with the bands giving rather unremarkable surface projections of the bulk NLs. In particular, the true shapes of the ``islands'' are given by their base boundaries i.e. intersection trenches exposed at low tide (tidal boundaries). This perspective suggests that it is the $\log|z|$ band intersections that are of decisive significance. Physically, this is indeed plausible: non-reciprocal similarity transforms can rescale\cite{yao2018edge,PhysRevLett.123.016805} $z=e^{ik_\perp}$, leading to ``tides'' or fluctuations of the sea level, but doing so will not affect the OBC spectrum which should be invariant under such basis transforms~\cite{PhysRevB.99.201103}.  As such, we call the topological surface states of non-Hermitian NKMs ``tidal'' states, in analogy to the well-known ``drumhead'' states that stretch across what we call the ``beaches''. 
Our formalism also trivially holds for Hermitian systems, in which the intersection trenches (tidal boundaries) are pinned to $\log |z|=0$, and hence coincide with the beaches.

To justify our marine analogy and explain how to choose the bands involved, we re-examine the criterion in Eq.~\ref{winding} in terms of the roots $z=a_i(\bold k_\parallel),b_i(\bold k_\parallel)$. It says ($\bold k_\parallel$ suppressed for brevity) that~\cite{PhysRevB.99.201103} a topological state exists at a given $\bold k_\parallel$ if the determinant set, i.e. set of the largest $r_a+r_b$ elements of $\{a_i\}\bigcup \{b_i\}$, does not contain $r_a$ elements from $\{a_i\}$ and $r_b$ elements from $\{b_i\}$. This implies the crucial role of $z_{r_a+r_b}(\bold k_\parallel)$, the $r_a+r_b$-th largest root in $\{a_i\}\bigcup \{b_i\}$, which gives the $r_a+r_b$-th highest $\log|z|$ band.

\noindent Consider a Trefoil knot NKM ($(p,q)=(2,3)$) with a $\hat n=\hat k_2$ surface termination. Suppose that $z_{r_a+r_b}\in \{a_i\}$ and the topological criterion is not satisfied, i.e., it belongs to a blue band in the $\log|z|$ band plots (Fig.~\ref{fig:XX2}d), which correponds to a point within the light blue region in the $k_3$-$k_1$ space (Fig.~\ref{fig:XX2}d) along the dotted line. In this case, there are $r_a(r_b)$ blue(red) bands in the determinant set. As $\bold k_\parallel$ varies, one of the following can happen to the $z_{r_a+r_b}$ band (colored blue): 
\begin{enumerate}[label=(\roman*)]
\item it intersects with another band in the determinant set; 
\item it intersects with another blue ($a_i$) band outside the determinant set, or 
\item it intersects with a red ($b_i$) band outside the determinant set. 
\end{enumerate}
Only for (iii) can one transit into the topological region, where the determinant set no longer consists of $r_a(r_b)$ blue (red) bands, as delineated by the interfaces between the light and dark colored regions in Fig.~{\ref{fig:XX2}c bottom panel.
Since the determinant set of the Trefoil knot~\cite{Methods} consists of the highest $r_a+r_b=4$ bands, the 4th (yellow) and 5th (brown) band in Fig.~{\ref{fig:XX2} uniquely determine its topological modes. These bands are plotted with $\hat e_1$ and $\hat e_2$ surface terminations in Fig.\ref{TrefoilBandStructure}(a-b). In more complicated NKMs where (ii) shows up, care must be taken in distinguishing the blue/red-type bands (Fig.~{\ref{fig:XX2}d) in the marine landscape of Fig.~{\ref{fig:XX2}c. While the above arguments for mapping out the topological tidal region may seem complicated, the alternative approach of manually keeping track of all the possible topological and non-topological skin gap closures of a 3D system is arguably more cumbersome. 

\begin{figure}
\centering
\includegraphics[width=\linewidth]{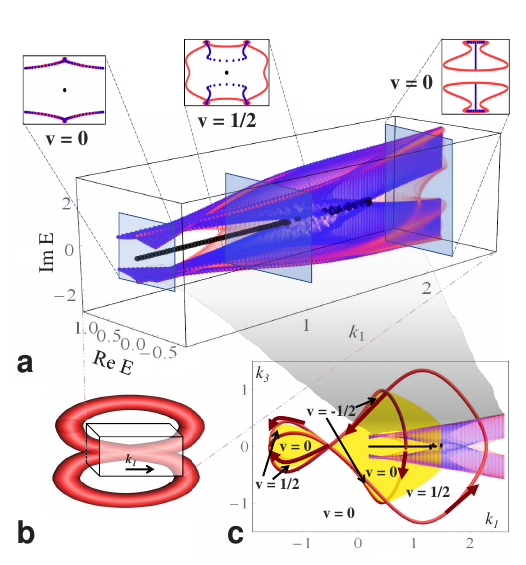}
\caption{Schematic illustration to relate the complex band structure, topological states, and  vorticity.}
\caption*{
(a) For the non-Hermitian Trefoil nodal knot metal (NKM), 3D plot of complex energy spectrum along $k_1$ and with $\hat{n}=\hat{e}_2$ surface, under periodic boundary conditions (PBC) and open boundary conditions (OBC) in red and blue respectively. The vorticity $v$ changes from 0 to $1/2$ and then $0$, as seen in the cross-section insets at $k_1=0.2,1$ and $2.2$. PBC states form red tubes enclosing the two blue OBC skin state branches, which can only meet and eliminate the topological modes (black) when the vorticity $v=1/2$. (b) The 3D PBC spectral plot is a segment of a Riemann surface (Bottom Left) obtained by closing the $k_1$ loop, with each pair of `pants' a vorticity transition. (c) The topological/tidal region (yellow) boundaries occur where the topological modes disappear and thus must lie in the regions with vorticity $v=1/2$, which experience a net anticlockwise winding of $\bold u$ (arrows).
}
\label{fig:XX3}
\end{figure}

\subsection{Relation to vorticity and skin states } We next highlight a special bulk-boundary correspondence between the OBC tidal region shape, which is subject to the skin effect along a specified boundary, and the conditions for their existence, which turns out to be related to the bulk point-gap topology i.e. vorticity. Recall that the tidal regions are purely determined by the imaginary gap band intersections (trenches in Fig.~{\ref{fig:XX2}a,c). However, the existence of these intersections is instead constrained by the PBC bulk nodal loop (NL) projections (dashed blue ``beaches" in Fig.~{\ref{fig:XX2}c). This is because imaginary gap bands intersect when the OBC skin gap (not PBC gap) closes~\cite{Methods}.  
As shown in the spectral inset plots of Fig.~\ref{fig:XX3}a, however, skin states (blue) generically accumulate in the interior of the PBC spectral loops (red)~\cite{PhysRevB.99.201103}. As such, skin gap closures can occur only when the skin states are contained within the same single PBC loop. This is just the condition of half-integer vorticity $v(\bold k_\parallel)=(\Gamma_a(1)+\Gamma_b(1))/(4\pi)$, which implies a branch cut in the energy Riemann surface~\cite{shen2018topological}. In terms of the bulk NLs, the vorticity at a point is half the number of times it is encircled anti-clockwise by the NL director $\bold u(\bold k)=\nabla_\bold k\, a(\bold k)\times \nabla_\bold k\, b(\bold k)$ along the surface-projected NL (Fig.~\ref{fig:XX3}c). 
Nontrivial vorticity does not obligate the skin states to intersect and further modify the determinant set makeup; whether this occurs depends on the $\log|z|$ band crossing intricacies (Fig.~\ref{fig:XX2}b).

\begin{figure}
\centering
\includegraphics[width=\linewidth]{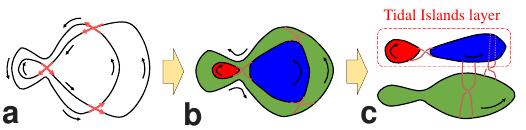}
\caption{Tidal islands from the Seifert surface of the dual nodal knot metal (NKM). }
\caption*{(a) The previously discussed dual Trefoil NKM with a reversed NL, such that each crossing has a reversed director $\bold u (\bold k)$. (b) The dual Seifert surface is constructed by promoting each crossing into a twist that connects regions bounded by the dual NL. (c) Resultant Seifert surface with a layer of islands isomorphic to the original tidal islands of vanishing vorticity. }
\label{fig:seifert}
\end{figure}

The vorticity argument laid out above can be visualized along any chosen path in the surface BZ (3D plot in Fig.~\ref{fig:XX3}a), where the PBC loci (red surface) becomes a Cobordism of one or more conjoined tube/s along the path, flanked by an interior skeleton (blue surface) of skin states. Within the tidal region (yellow) in Fig.~\ref{fig:XX3}c, topological modes also exist as additional isolated strands (black). The tubes of a closed path will be joined at their ends, forming a Riemann surface (red) indicative of the vorticity structure (Fig.~\ref{fig:XX3}b). For the 2-band model we studied, there are at most two parallel tubes (PBC bands). In generic multi-band cases, far more interesting Riemann surfaces can be obtained, where each ``pair of pants'' in its decomposition corresponds to a vorticity transition. 
Equivalently, the tidal boundaries, being $\log|z|$ band crossings, can also be viewed as trajectories of the surface projected NL crossings under complex analytic continuation $k_\perp\rightarrow k_\perp-i\log z$. As such, from non-Hermitian tidal regions, we gain access to the band structure in the complex momentum domain, and not just the real domain as from Hermitian drumhead regions.

\subsection{Tidal states and their dual Seifert surfaces } 
We highlight another interesting link between the topological configuration of the tidal regions and the Seifert surface of its dual NKM. As a surface bounding a knot, a Seifert surface contains various useful information about the knot topology i.e. the Alexander knot invariant polynomial can be extracted from its homology generators~\cite{murasugi2007knot,li2018realistic,li2019emergence}. As an extended object, it can be experimentally detected more easily than the nodal structure itself too~\cite{li2019emergence}.

To proceed, recall that the vorticity determines the topology (connectedness) of the tidal region shape, and is deeply related with topological invariants of the nodal knot. 
As previously explained, tidal boundaries cannot penetrate regions of zero vorticity. The tidal regions are hence topologically constrained to contain islands of vanishing vorticity. To endow these islands with further topological significance, we appropriately reverse the directors $\bold u(\bold k)$ of certain NLs such that each crossing in the knot diagram has a reversed director (compare Figs.~\ref{fig:XX3}c and Fig.~\ref{fig:seifert}a). This defines a ``dual" NKM which bounds a Seifert surface~\cite{murasugi2007knot} that, strikingly, exhibits a 
layer structure resembling our tidal islands (Fig.~\ref{fig:seifert}(a-c)). Fig.~\ref{fig:seifert}b is a sample construction of the dual $\hat e_2$ Seifert surface of the beforementioned Trefoil NKM, from which the islands of zero vorticities metamorphosize into two disconnected Seifert surface regions isomorphic to the original tidal islands.
Intricate relations exist between these islands and NKM topology. For NKMs embedded in $\mathbb{R}^3$, the surface projection of a dual NKM with $C$ crossings, $L$ NLs and $X$ disconnected tidal regions yields a genus $G=(1+C-X-L)/2$ dual Seifert surface with $2G+L-1$ homology generators~\cite{murasugi2007knot}. Distinct from the Fermi surface realizations discussed in Ref.~\onlinecite{carlstrom2018exceptional}, our dual Seifert surfaces also contains topological information through the linking matrix $S$ of its homology generators~\cite{murasugi2007knot}.  Specifically, knot invariants such as the Alexander polynomial and the knot signature are respectively given by $A(t)=t^{-G}\text{Det}(S-tS^T)$ and $\text{Sig}(S+S^T)$. 

\begin{figure}
\centering
\includegraphics[width=\linewidth]{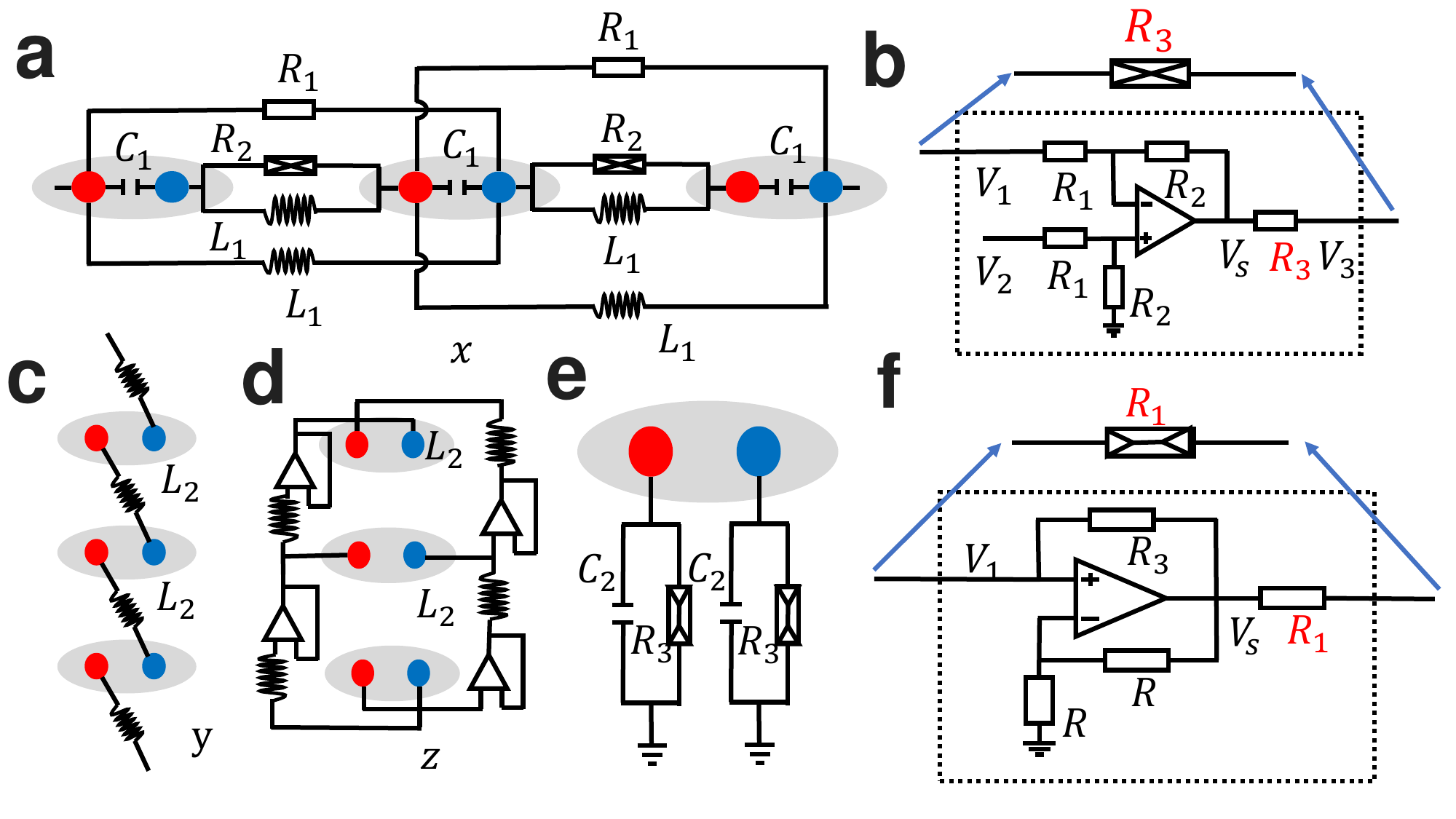}
\caption{Illustration of the various constituents of the Hopf link nodal knot metal (NKM) circuit.}
\caption*{(a,c,d) Nearest neighbor (NN) hopping in the (a) $x$ direction,  (c) $y$ direction, (d) $z$ direction. (e) Extra grounded (on-site) hopping to cancel the diagonal terms in the Laplacian. (b,f) Non-reciprocal feedback for skin (tidal) effects is implemented via (b) a differential amplifier and (f) a negative resistor.}
\label{fig:circuit}
\end{figure}

\subsection{Discussion } Non-Hermitian nodal knot metals (NKMs) reach far beyond their Hermitian counterparts in terms of conceptual significance and even potentially allow for more practical realizations. We demonstrated this using the Hopf-Link and Trefoil knot NKM (the detailed complex spectra in Figs.~\ref{Hopfe1},\ref{Hopfe2},\ref{Hopfe3},\ref{Trefoile1},\ref{Trefoile2},\ref{Trefoile3}). Equipped with a generalized recipe for constructing non-Hermitian NKMs with unprecedentedly short-ranged hoppings, we reveal the algebraic, geometric, and topological aspects of their topological surface states via a marine analogy formalism, where ``tidal" intersection boundaries beneath the $\log|z|=0$ Bloch sea are identified as pivotal in defining topological phase boundaries. While the tidal region geometry depends on algebraic quantities such as the imaginary gap crossings, its topology depends, via complex band vorticity, on not just knot topology, but also orientation. A dual Seifert surface interpretation uncovers this new link between knot topology and the tidal islands, thereby helping to bridge the seeming conceptual disconnect between band structure and eigenstate topology.

The NKMs discussed in this work can be straightforwardly realized in electrical circuit Laplacian bandstructures. Instead of the Hamiltonian, it is the Laplacian $J$ that determines the steady state behavior of a circuit via $\bold I = J\bold V$, where $\bold I$ and $\bold V$ are the input currents and potentials at all the nodes. Since the circuit engineering of desired Laplacian is a mature topic~\cite{ningyuan2015time,Lee2018,Imhof2018,PhysRevB.99.161114} with Hermitian nodal drumhead states even experimentally measured~\cite{gao2018experimental,deng2019nodal,lee2020imaging}, we shall relegate its details to the Supplement~\cite{Methods}. As illustrated in Fig.~\ref{fig:circuit}(a-f), the key point is that positive, negative and non-reciprocal couplings can be simulated with appropriate combinations of RLC components and operation amplifiers (op-amps), which introduces non-reciprocal feedback needed for the (tidal) skin effect. Upon setting up the circuit, its Laplacian and bandstructure can be reconstructed through systematic impedance and voltage measurements between each node and the ground~\cite{PhysRevB.99.161114}. In particular, topological zero modes reveal themselves through divergent impedances known as topolectrical resonances, thus allowing tidal regions to be mapped out as parameter regions of very large impedances~\cite{Methods}.

\section{Methods}

\subsection{Construction of non-Hermitian nodal knot metals (NKMs)}

\noindent We look at 2-component models of the form
\begin{equation}
H(\bold k)=h_x(\bold k)\sigma_x+h_y(\bold k)\sigma_y+h_z(\bold k)\sigma_z=\bold h(\bold k)\cdot \mathbf{ \sigma},
\label{Hxyz_supp}
\end{equation}
where $\bold k\in \mathbb{T}^3$ and $\sigma_{x,y,z}$ are the Pauli matrices. The gap in such a minimal model is proportional to 
\begin{equation}
f(\bold k)=h_x(\bold k)^2+h_y(\bold k)^2+h_z(\bold k)^2.
\label{fh}
\end{equation}
Hence the engineering of $\bold h(\bold k)$ for realizing certain desired non-Hermitian nodal knots or links is broken down into two tasks: (i) finding the appropriate $f(\bold k)$ that vanishes along the desired knot/link trajectory and (ii) choosing $\bold h(\bold k)$ components that approximately but adequately satisfies Eq.~\ref{fh}.

\begin{comment}
In general, $(p,q)$-NKMs can be generated by choosing $\bold h(\bold k)$ such that~\cite{bode2017knotted,ezawa2017topological,Methods}
\begin{equation}
h_x^2+h_y^2+h_z^2 = \mu^p+w^q=f(\mu,w),
\label{hxyz3}
\end{equation}
where $(\mu(\bold k),w(\bold k))$ maps the 3D BZ $\mathbb{T}^3$ onto $\mathbb{C}^2$ with nontrivial winding number. Choosing them to be the regularized stereographic projection, we have
\begin{eqnarray}
\mu(\bold k)&=&\sin k_3 + i(\cos k_1 +\cos k_2+\cos k_3-m),\qquad\notag\\
w(\bold k)&=&\sin k_1 + i\sin k_2, \qquad \quad \,\,\, 1.5<m<2.5.\qquad
\end{eqnarray}
More generally, $\bold h(\bold k)$ of a large class of non-Hermitian NKMs can be similarly approximated by the ``square root" of their Hermitian counterparts.
\end{comment}

\subsubsection{Knots from braids}

To obtain a possible form for $f(\bold k)$, we first review how the Hermitian case has been handled. In one intuitive approach, the knot/link is first defined as a braid closure, which is then ``curled'' up in the 3D BZ 
. Consider a braid with $N$ strands taking 
complex position coordinates $\mu_1(t),\mu_2(t),...,\mu_N(t)$, where $t$ is the ``time" parametrizing the braiding processes. Since the ends of the strands are joined to form the knot/link, we compactify $t\rightarrow e^{it}$ and introduce a braiding function 
\begin{equation}
\bar f(\mu,e^{it})= \prod_j^N \left(\mu-\mu_j(t)\right)
\label{fzw}
\end{equation}
such that $\bar f(\mu,e^{it})=0$ is precisely satisfied along the braids. 
To appropriately ``curl'' up the braid into the 3D BZ, we next analytically continue $\bar f(\mu,e^{it})$ into $f(\mu(\bold k),w(\bold k))$, where $\mu=\mu(\bold k)$ and $w=w(\bold k)$ are two complex functions of the momentum $\bold k$ in the BZ. The kernel of $f(\mu(\bold k),w(\bold k))=0$ then gives the knot/link in the BZ, which can be implemented as a nodal structure. 
Inspired by the stereographic projection, we shall choose $\mu(\bold k),w(\bold k)$ to be its regularized form~\cite{PhysRevB.99.161115}:
\begin{eqnarray}
\mu(\bold k)&=&\sin k_3 + i(\cos k_1 +\cos k_2+\cos k_3-m),\qquad\notag\\
w(\bold k)&=&\sin k_1 + i\sin k_2, \qquad \quad \,\,\, 1.5<m<2.5,\qquad
\label{zw}
\end{eqnarray}
which faithfully maps the braid closures into the 3D BZ, as attested by its winding number\cite{bi2017nodal} from $\mathbb{T}^3$ to $\mathbb{C}^2$. The value of $m$ is chosen such that it does not introduce any extraneous nodal structures in the BZ. By considering their braids, it can be shown that $f(\mu(\bold k),w(\bold k))=z(\bold k)^p+w(\bold k)^q$ for generic $(p,q)$-torus knots.

As a simplest illustration, consider the Hopf link NKM, which is formed by closing $N=2$ strands parametrized by $\mu_j (s)=i(-1)^je^{is}$. Its braiding function is $\bar f(\mu,e^{is})=(\mu-ie^{is})(\mu+ie^{is})$, which yields $f(\mu,w)=\mu^2+w^2=0$ along the link. Since $\mu(\bold k),w(\bold k)$ are complex (Eq.~\ref{zw}), they can only directly enter the components of $\bold h(\bold k)$ in the non-Hermitian case. As such, a possible realization for the non-Hermitian Hopf link is $\bold h(\bold k)=(\mu(\bold k),w(\bold k),0)$, which contains only nearest-neighbor hoppings. But contrast, the Hermitian case requires a more complicated $\bold h(\bold k)$ that contains $\text{Re}\, f$ and $\text{Im}\, f$, which also includes next-nearest-neighbor hoppings (2nd Fourier coefficients in $\bold k$).  

By Fourier expanding $H(\bold k)$ of the non-Hermitian Hopf link, we obtain its real-space hopping coefficients illustrated in Fig.~\ref{fig:braid}(a-d) of the main text. Specifically, in the $k_3=0$ plane,
\begin{eqnarray}
&&H^\text{non-Herm}_{12}(\bold k)= -i + \frac{1}{2}(1+i)e^{-i k_1} - \frac{1}{2}(1-i)e^{i k_1} + i e^{-i k_2}\notag \\
&&H^\text{non-Herm}_{21}(\bold k)= -i + \frac{1}{2}(1+i)e^{i k_1} - \frac{1}{2}(1-i)e^{-i k_1} + i e^{i k_2} \notag\\
&&H^\text{non-Herm}_{11}(\bold k)=H^\text{non-Herm}_{22}(\bold k)=0,
\end{eqnarray}
which gives for instance a hopping of $-i$ between the two sites of the same sublattice, and a complex hoppings of $\pm\frac{1\pm i}{2}$ between different sublattice sites of adjacent unit cells separated by $\hat e_1$. These nearest-neighbor hoppings are to be contrasted with further next-nearest-neighbor hoppings of the corresponding Hermitian Hopf Hamiltonian with $h_x=\text{Re}(z^2+w^2)$, $h_y=\text{Im}(z^2+w^2)$ and $h_z=0$. In the $k_3=0$ plane, we have
\begin{eqnarray}
H^\text{Herm}_{12}&=& \frac{1}{2}\bigg[-4 + 2 (e^{-i k_1}+e^{i k_1}+e^{-i k_2}+e^{i k_2}) \nonumber \\&&-(e^{-2 i k_1}+e^{2 i k_1}) 
-(1+i)(e^{-i(k_1-k_2)}+e^{i(k_1-k_2)}) \nonumber \\&&- (1-i)(e^{-i(k_1+k_2)}+e^{i(k_1+k_2)})\bigg]\nonumber \\
H^\text{Herm}_{21}&&=H^{*\text{Herm}}_{12}\\
H^\text{Herm}_{11}&&=H^\text{Herm}_{22}=0
\end{eqnarray}
which is also illustrated in Fig.~\ref{fig:braid}a of the main text.

\subsubsection{Explicit ansatz for $(p,q)$-torus knots}

We now derive Eq.~5 of the main text, which a rather generic ansatz for torus knots:
\begin{equation}
\bold h=\left\{\begin{array}{lll}
(2\mu^i,2w^j,0),& &p=2i,q=2j\\
(2\mu^i,w^j+w^{j+1},\gamma),& &p=2i,q=2j+1\\
(\mu^i+\mu^{i+1},w^j+w^{j+1},\gamma),& &p=2i+1,q=2j+1,\qquad
\end{array}\right.
\label{hpq_supp}
\end{equation}
For $p=2i$, $q=2j$, it is obvious that $f=4(\mu^p+w^q)$ vanishes exactly at where we wanted. The situation is more tricky when either $p$ or $q$ is odd. Suppose $q=2j+1$ is odd, in this case, we cannot simply take the square root of $w^{2j+1}$, since that will contain non-integer powers of the trigonometric functions of $\bold k$ components. In our ansatz, we replace $2w^{(2j+1)/2}$ by $w^j+w^{j+1}$, which amounts to replacing the geometric mean of $w^j$ and $w^{j+1}$ with their arithmetic mean, and also add a $h_z=\gamma$ component. To understand the role of $\gamma$, suppose for a moment that it is omitted. Doing so, we have an unwanted degeneracy ($f=0$) at $\bold k=(-\pi/2,0,0)$ for $m=2$, which corresponds to $w=-1$ and $\mu=i(2-m)=0$. To lift this degeneracy, we either perturb $m$ away from $2$, or is forced to introduce a nonzero $\gamma$. It turns out that the latter option gives us more consistent control over a large number of possible $p$ and $q$. For the Trefoil ($p=2,q=3$) knot for instance, a real $\gamma$ breaks the degeneracy and gives a Hopf link, while an imaginary $\gamma$ gives the desired Trefoil knot. 

In Hermitian NKMs, $\bold h(\bold k)$ is real, and Eq.~\ref{fzw} can only be satisfied by letting its nonzero components be $\text{Re}\, f$ or $\text{Im}\, f$. A key simplification occurs, however, for non-Hermitian NKMs where $\bold h(\bold k)$, being complex, can actually take simpler forms. This insight does not emerge if one intends to obtain non-Hermitian knots/links just by perturbing known Hermitian nodal structures~\cite{PhysRevB.99.081102,PhysRevB.99.075130,carlstrom2018exceptional}. For illustration, the simplest non-Hermitian Hopf-link NKM ($(p,q)=(2,2)$) can be generated with $\bold h(\bold k)=(\mu(\bold k),w(\bold k),0)$, with $|\bold h(\bold k)|$ the square root of $\bold h(\bold k)=(\text{Re}\,f,\text{Im}\,f,0)$, $f=\mu(\bold k)^2+w(\bold k)^2$. The Hermitian Hopf NKM thus necessitates twice the coupling range in comparison to its non-Hermitian analog.

\subsection{Tidal regions and their relation to vorticity}

In Fig.~\ref{fig:vorticity}(a-f), we show the surface state plots of various torus knots, some with more than one type of surface termination. The topological (tidal) surface state regions (translucent red) are superimposed onto the vorticity regions (green and cyan), clearly demonstrating that tidal boundaries are totally contained within regions of nonzero vorticity, a necessary condition for the gap closure of the skin states. As discussed in the main text, the collorary is that the tidal state islands must therefore surround islands of zero vorticity (uncolored), which will be evident below.

\begin{figure}
\centering
\includegraphics[width=\linewidth]{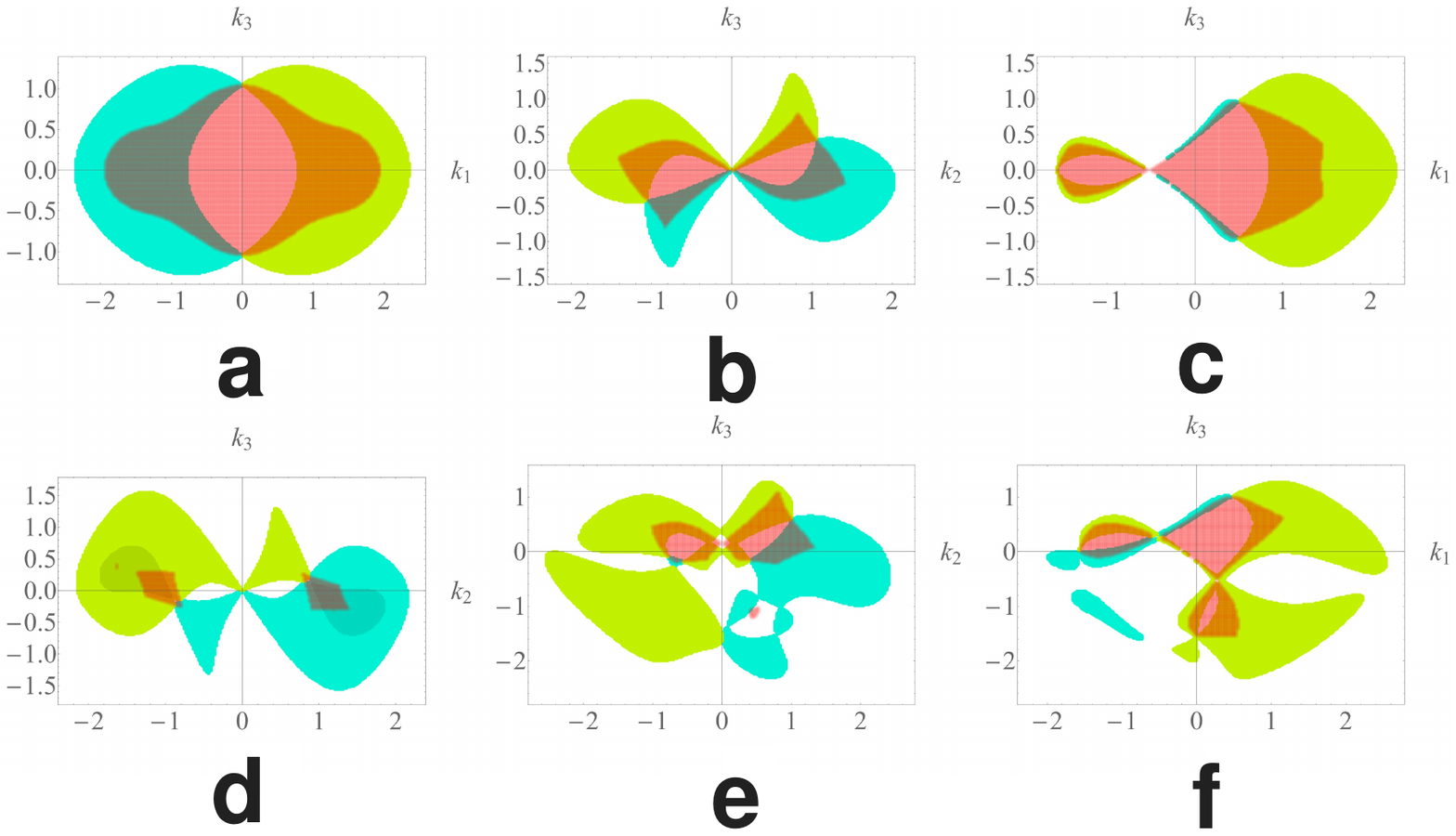}
\caption{Tidal regions (translucent red) superimposed onto regions of different vorticities $v=1,\frac1{2},-\frac1{2}$ and $-1$, colored dark green, green, cyan and dark cyan respectively.}
\caption*{The respective nodal knot metals (which are $(p,q)$-torus knots) with their surface termination normals $\hat{e}_n$: (a) $\hat e_2$ Hopf-link ($p=q=2$), (b) $\hat e_1$ Trefoil ($p=2,q=3$), (c) $\hat e_2$ Trefoil, (d) $\hat e_1$ ($p=2,q=5$), (e) $\hat e_1$ 3-link ($p=3,q=3$) and (f) $\hat e_2$ 3-link. In all cases, the tidal islands surround a region of zero vorticity (white). In (d), this white region is very tiny, lying at the intersection of $\pm \frac{1}{2}$ vorticity regions. Note that boundaries between vorticities of the same sign are ignored, since the skin states can intersect as long as the vorticity does not change sign at a $v=0$ point.
}
\label{fig:vorticity}
\end{figure}

\newpage

\begin{widetext}
\subsection{Details of circuit realizations of non-Hermitian Hopf and Trefoil knot nodal circuits}

Here, we provide explicit details of the construction of circuit Laplacians $J$ with NKM bandstructures. We use circuits with 2 sites per unit cells, in other to form two band models $J=J_{\alpha\beta}$, $\alpha,\beta=1,2$. 

\subsection{Hopf-link - a (2,2)-torus knot}
From Eqs.~\ref{Hxyz},\ref{hpq},\ref{muw} of the main text with the Hamiltonian replaced by $J$, we have for the Non-Hermitian Hopf Hamiltonian $d_x=d_{1x}+\mi~d_{2x}=z$, $d_y=d_{1y}+\mi~d_{2y}=w$ and $d_z=d_{1z}+\mi~d_{2z}=0$,
\begin{eqnarray}\label{eq:nonhermhopf}
&&J_{11}=0, \ \ J_{12}=z-\mi~w =-2\mi  + \frac{1}{2}(\mi + 1)\me^{-\mi k_1} + \frac{1}{2}(\mi - 1)\me^{\mi k_1} + \mi \me^{-\mi k_2} + \mi \me^{-\mi k_3} ;\nonumber \\
&&J_{21}=z+\mi~w = -2\mi + \frac{1}{2}(\mi+1)\me^{\mi k_1} + \frac{1}{2}(\mi - 1)\me^{-\mi k_1} + \mi \me^{\mi k_2}+\mi \me^{-\mi k_3}, \ \ J_{22}=0.
\end{eqnarray}
Because we only have off-diagonal terms $J_{12}$ and $J_{21}$, these two terms should have the form 
\begin{eqnarray}\label{eq:Hoff}
J_{\mathrm{off}}= - \left( \sum_{i} \mi \omega C_i + \sum_{k} \frac{1}{\mi \omega L_{k}} + \sum_{j}\frac{1}{R_j} \right) =- \mi \omega \bigg( \sum_{i}  C_i  - \sum_{k} \frac{1}{\omega^2 L_{k}}- \mi \sum_{j}\frac{1}{\omega R_j}\bigg) .
\end{eqnarray}
Here the challenge we are facing is that we only have real positive(capacitors), real negative(inductors) and imaginary negative(resistors) terms. However, in $J_{12}$ and $J_{21}$, there are imaginary positive terms. This challenge still persists even we take $\mi$ or $-\mi$ out of the total Hamiltonian. There are two possible ways out: (i) modify the Hamiltonian such that it contains no imaginary positive terms or equivalent forms, or (ii) try to construct imaginary positive terms through active circuit elements.

\subsubsection{Imaginary positive terms using differential amplifier}
First we introduce differential amplifier (Fig.~\ref{fig:nonHermelement}a), the current $I_1$, $I_2$ and $I_3$ flows from input $V_1$, $V_2$ and $V_3$. 
\begin{eqnarray}
&&V_1-I_1 R_1 =V_2-I_2 R_1, \ \ V_s + I_1 R_2 = I_2 R_2, \ \ I_3=\frac{1}{R_3}(V_3-V_s) \\
&&\Rightarrow I_3 =\frac{1}{R_3} V_3 - \frac{R_2}{R_1R_3} V_2 +\frac{R_2}{R_1R_3} V_1
\end{eqnarray}
If we connect $V_2$ to the ground, and $R_1=R_2$, we will have $I_3=\frac{1}{R_3}(V_3+V_1)$. If we consider $V_1$ as input and $V_3$ as output, we will have a positive term $\frac{1}{R_3}=-\mi\omega (\mi \frac{1}{\omega R_3})$, which is an imaginary positive term added to Eq.(\ref{eq:Hoff}).

\begin{figure}
\centering
\includegraphics{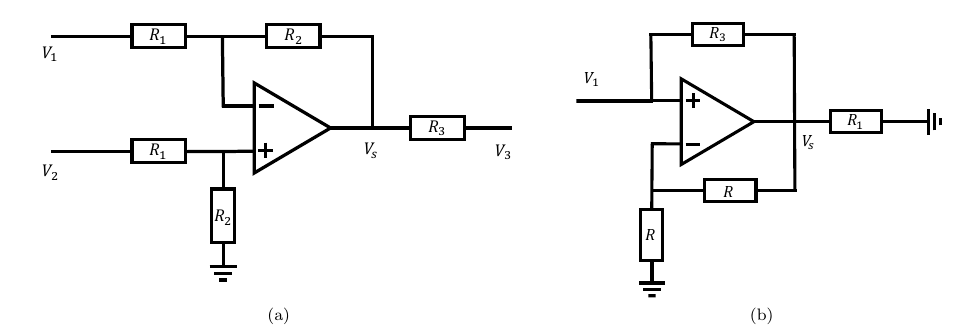}
\caption{(a) Differential amplifier and (b) negative resistor as active non-Hermitian elements.}
\label{fig:nonHermelement}
\end{figure}

\subsubsection{Non-Hermitian Hopf-link circuits}

We can write down the Laplacian of the Hopf link, whose circuit constituents are illustrated in Fig.~\ref{fig:circuit} of the main text:
\begin{eqnarray}
&&J_{11}=J_{22}= \mi \omega C_1 + \frac{2}{\mi\omega L_1} + \frac{2}{\mi\omega L_2} + \frac{1}{R_1} + \frac{1}{R_2}, \\
&&J_{12}=-\mi\omega C_1 - \frac{1}{R_1} \me^{\mi k_1} - \frac{1}{\mi \omega L_1}\me^{\mi k_1} + \frac{1}{R_2} \me^{-\mi k_1} - \frac{1}{\mi \omega L_1}\me^{-\mi k_1}  - \frac{1}{\mi \omega L_2}\me^{-\mi k_2}- \frac{1}{\mi \omega L_2}\me^{-\mi k_3} \nonumber \\
&&=\frac{1}{2} \omega C_1  \bigg[  -2\mi + \bigg(\frac{4}{\omega^2 C_1 L_1}\bigg) \frac{\mi}{2}\me^{-\mi k_1} + \bigg(\frac{4}{\omega C_1 R_2}\bigg) \frac{1}{2}\me^{-\mi k_1}+\bigg(\frac{4}{\omega^2 C_1 L_1}\bigg) \frac{\mi}{2}\me^{\mi k_1} - \bigg(\frac{4}{\omega C_1 R_1}\bigg) \frac{1}{2}\me^{\mi k_1}\nonumber \\
&& + \bigg(\frac{2}{\omega^2 C_1 L_2}\bigg) \mi \me^{-\mi k_2} + \bigg(\frac{2}{\omega^2 C_1 L_2}\bigg) \mi \me^{-\mi k_3}   \bigg]. \\
&&J_{12}=-\mi\omega C_1 - \frac{1}{R_1} \me^{-\mi k_1} - \frac{1}{\mi \omega L_1}\me^{-\mi k_1} + \frac{1}{R_2} \me^{\mi k_1} - \frac{1}{\mi \omega L_1}\me^{\mi k_1}  - \frac{1}{\mi \omega L_2}\me^{\mi k_2}- \frac{1}{\mi \omega L_2}\me^{-\mi k_3}\nonumber \\
&&=\frac{1}{2} \omega C_1  \bigg[  -2\mi + \bigg(\frac{4}{\omega^2 C_1 L_1}\bigg) \frac{\mi}{2}\me^{\mi k_1} + \bigg(\frac{4}{\omega C_1 R_2}\bigg) \frac{1}{2}\me^{\mi k_1}+\bigg(\frac{4}{\omega^2 C_1 L_1}\bigg) \frac{\mi}{2}\me^{-\mi k_1} - \bigg(\frac{4}{\omega C_1 R_1}\bigg) \frac{1}{2}\me^{-\mi k_1}\nonumber \\
&& + \bigg(\frac{2}{\omega^2 C_1 L_2}\bigg) \mi \me^{\mi k_2} + \bigg(\frac{2}{\omega^2 C_1 L_2}\bigg) \mi \me^{-\mi k_3}   \bigg].
\end{eqnarray}
As we can see, if we want to simulate Eq.(\ref{eq:nonhermhopf}) using circuits, we'd better have the following configurations of these elements,
\begin{eqnarray}
\omega^2 C_1 L_1 =4, \ \ \omega C_1 R_1 =\omega C_1 R_2=4, \ \ \omega C_1 L_2 =2,
\end{eqnarray}
which means that
\begin{eqnarray}
J_{11}=J_{22}=\frac{\mi \omega C_1}{2}(-i-1).
\end{eqnarray}
To cancel this term, we need to connect every red and blue point with a capacitor $C_2=C_1/2$ and a negative resistance $R_3$ as shown by Fig.~\ref{fig:nonHermelement}b. The negative resistance gives
\begin{eqnarray}
I_1 = \frac{V_1 -V_s}{R_3}, \ \ V_s=2 V_1 \Rightarrow I_1=-V_1/R_3, \ \ \omega C_1 R_3=2.
\end{eqnarray} 

\subsection{Trefoil knot - a (2,3)-torus knot}
\subsubsection{Non-Hermitian Trefoil knot circuits}
Similar to the non-Hermitian Hopf-link circuits, non-Hermitian Trefoil knot circuits can also be constructed using a combination of capacitors, inductors, resistors, differential amplifier and negative resistors. First of all, we write down the model proposed in the main text:
\begin{eqnarray}
&&J_{11}=-J_{22}=\frac{1}{2}\mi , \\
&&J_{12}=-2\mi + \bigg(\frac{\mi}{2}+\frac{1}{4}\bigg)\me^{-\mi k_1}+ \bigg(  \frac{\mi}{2} - \frac{1}{4}\bigg)\me^{\mi k_1} + \frac{1}{8}\mi \me^{-2\mi k_1} + \frac{1}{8}\mi \me^{2\mi k_1} - \frac{1}{4} \me^{-\mi k_1 - \mi k_2} - \frac{1}{4} \me^{\mi k_1 + \mi k_2}+\frac{1}{4} \me^{\mi k_1 - \mi k_2} + \frac{1}{4} \me^{-\mi k_1 + \mi k_2} \nonumber \\
&& + \frac{3}{4}\mi \me^{-\mi k_2} + \frac{1}{4}\mi \me^{\mi k_2} -\frac{1}{8}\mi \me^{-2\mi k_2} -\frac{1}{8}\mi \me^{2\mi k_2} + \mi \me^{-\mi k_3}, \\
&&J_{21}=-2\mi + \bigg(\frac{\mi}{2}+\frac{1}{4}\bigg )\me^{\mi k_1}+ \bigg(  \frac{\mi}{2} - \frac{1}{4}\bigg)\me^{-\mi k_1} - \frac{1}{8}\mi \me^{-2\mi k_1} - \frac{1}{8}\mi \me^{2\mi k_1} + \frac{1}{4} \me^{-\mi k_1 - \mi k_2} + \frac{1}{4} \me^{\mi k_1 + \mi k_2}-\frac{1}{4} \me^{\mi k_1 - \mi k_2} - \frac{1}{4} \me^{-\mi k_1 + \mi k_2} \nonumber \\
&& + \frac{3}{4}\mi \me^{\mi k_2} + \frac{1}{4}\mi \me^{-\mi k_2} +\frac{1}{8}\mi \me^{-2\mi k_2} +\frac{1}{8}\mi \me^{2\mi k_2} + \mi \me^{-\mi k_3}.
\end{eqnarray}
\begin{figure}[t]
\includegraphics[width=0.9 \linewidth]{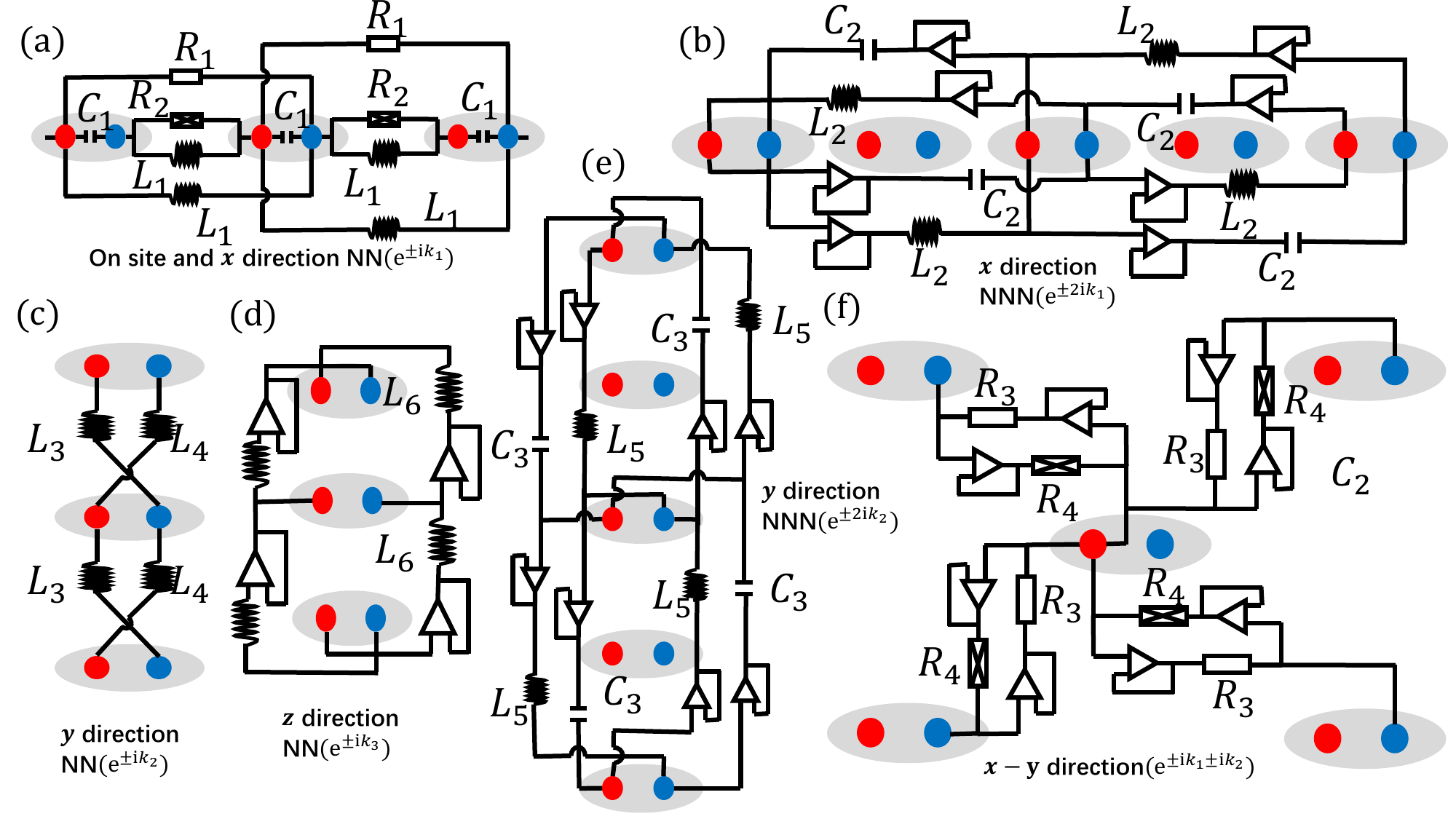}
\caption{Constituents of the non-Hermitian Trefoil knot circuit}
\caption*{(a) On-site hopping and nearest neighbour (NN) hopping in the $x$ direction, (b) next nearest neighbour (NNN) hopping in the $x$ direction, (c) NN  hopping in the $y$ direction, (d) NN hopping in the $z$ direction, (e) NNN hopping in the $y$ direction and (f) diagonal hopping in the $x-y$ plane.  }
\label{fig:trefoilcircuits}
\end{figure}
According to Fig.~\ref{fig:trefoilcircuits}, we can write down its Laplacian,
\begin{eqnarray}
&&J_{11} = \mi \omega C_1 + \frac{1}{R_1} + \frac{1}{R_2} + \frac{2}{\mi\omega L_1} + \frac{2}{\mi\omega L_2} + \frac{1}{\mi \omega L_3} +\frac{1}{\mi\omega L_4} +\frac{1}{\mi\omega L_6} + 2\mi \omega C_3 + \frac{2}{R_3}+\frac{2}{R_4}, \\
&&J_{12}=-\mi \omega C_1 - \frac{1}{R_1}\me^{\mi k_1} + \frac{1}{R_2}\me^{-\mi k_1} - \frac{1}{\mi\omega L_1}(\me^{-\mi k_1}+\me^{\mi k_1}) -\frac{1}{\mi\omega L_2}(\me^{-2\mi k_1}+\me^{2\mi k_1}) -\frac{1}{\mi \omega L_3}\me^{-\mi k_2} -\frac{1}{\mi\omega L_4}\me^{\mi k_2} - \frac{1}{\mi \omega L_6}\me^{-\mi k_3} \nonumber \\
&&- \mi \omega C_3(\me^{-2\mi k_2}+\me^{2\mi k_2}) + \frac{1}{R_4}(\me^{-\mi k_1+\mi k_2} + \me^{\mi k_1-\mi k_2}) -\frac{1}{R_3}(\me^{-\mi k_1-\mi k_2} + \me^{\mi k_1+\mi k_2}), \\
&& J_{21}=-\mi \omega C_1 - \frac{1}{R_1}\me^{-\mi k_1} + \frac{1}{R_2}\me^{\mi k_1} - \frac{1}{\mi\omega L_1}(\me^{-\mi k_1}+\me^{\mi k_1}) -\mi\omega C_2(\me^{-2\mi k_1}+\me^{2\mi k_1}) -\frac{1}{\mi \omega L_3}\me^{\mi k_2} -\frac{1}{\mi\omega L_4}\me^{-\mi k_2} - \frac{1}{\mi \omega L_6}\me^{-\mi k_3} \nonumber \\
&&-\frac{1}{ \mi \omega L_5}(\me^{-2\mi k_2}+\me^{2\mi k_2}) - \frac{1}{R_3}(\me^{-\mi k_1+\mi k_2} + \me^{\mi k_1-\mi k_2}) +\frac{1}{R_4}(\me^{-\mi k_1-\mi k_2} + \me^{\mi k_1+\mi k_2}), \\
&& J_{22}= \mi \omega C_1 + \frac{1}{R_1} + \frac{1}{R_2} + \frac{2}{\mi\omega L_1} + 2\mi\omega C_2 + \frac{1}{\mi \omega L_3} +\frac{1}{\mi\omega L_4} +\frac{1}{\mi\omega L_6} + \frac{2}{\mi \omega L_5} + \frac{2}{R_3}+\frac{2}{R_4}.
\end{eqnarray}
Comparing these two Hamiltonians from toy model and circuits, we can find 
\begin{eqnarray}
&&\omega C_1 R_1=\omega C_1 R_2= \omega C_1 R_3=\omega C_1 R_4=8, \ \ \omega^2 C_1 L_1=4, \ \ \omega^2 C_1 L_2=\omega^2 C_1 L_5=16, \nonumber \\
&& \omega^2 C_1 L_3 = \frac{8}{3},\ \ \omega^2 C_1 L_4=8, C_2=C_3=\frac{1}{16}C_1, \ \ \omega^2 C_1 L_6 =2.
\end{eqnarray} 
which means that
\begin{eqnarray}
J_{11}=J_{22}=\frac{1}{2}\omega C_1 \bigg(\frac{3}{2}-\mi\bigg).
\end{eqnarray}
From this, we should connect every red and blue point with a capacitor $C_4=(C_1+1/\omega)/2$ and a negative resistor $R_5=\frac{4}{3\omega C_1}$. 

\begin{figure}[H]
\centering
\includegraphics[width=0.6\linewidth]{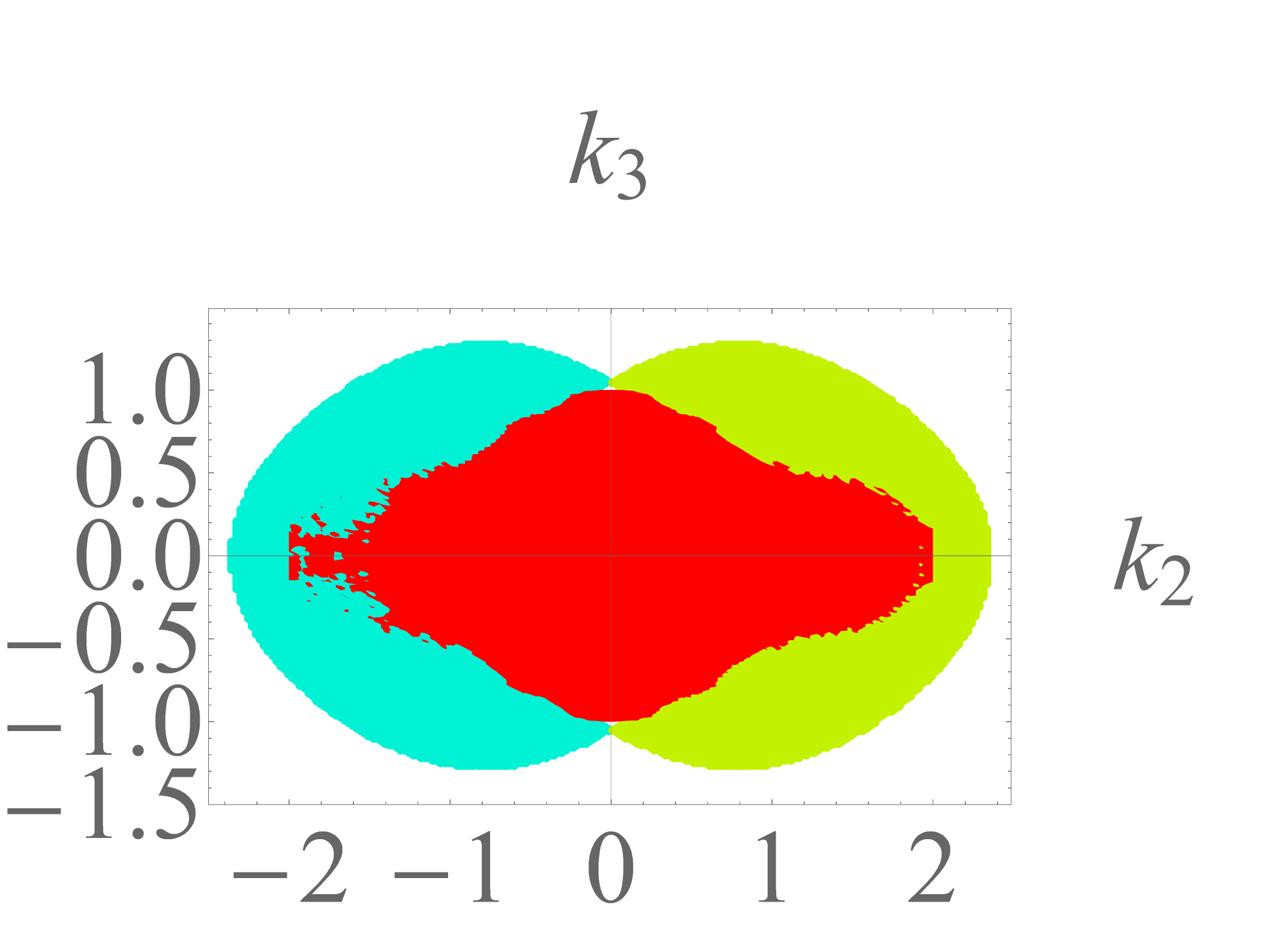}
\caption{Region of divergent node-averaged impedance (red) in the parameter space of a simulated Hopf circuit, which agrees very well with the tidal region of the Hopf link nodal knot metal (NKM). The cyan and green regions are the surface projections of the bulk nodal structure (conventional drumhead regions), which are totally different from the tidal regions. }
\label{fig:hopfresonances}
\end{figure}
\subsubsection{Topolectrical Resonance for mapping of topological zero modes}

As is well-known~\cite{Lee2018}, the presence of a zero eigenvalue in the Laplacian causes divergent impedances, as is seen through $\bold V=J^{-1}\bold I$. This extreme sensitivity of $\bold V$ with $\bold I$, limited above by parasitic and contact resistances, is known as topolectrical resonance. Fig.~\ref{fig:hopfresonances} shows the region of divergent resistance of the Hopf link circuit in parameter space, averaged out over random pairs of nodes. Not surprisingly, the region of topolectrical resonance agrees excellently with the theoretically predicted tidal region.

\end{widetext}
\section*{Data Availability}
The data that support the plots within this paper and other findings of this study are available from any of the authors upon reasonable request.

\section*{Code Availability}
Though not essential to the central conclusions of this work, computer codes used to generate Figs. 2-4 and 6-9 in the main text, and Supplementary Figs. S1–7 are available upon reasonable request via email to C.H.L.

\section*{Author Contributions}
CH Lee, R Thomale and X Zhang initiated the project. CH Lee conceived the ideas presented, and performed the initial computations. X Zhang, G Li, Y Liu, T Tai contributed to the results of the topological states of the non-Hermitian Hopf link and Trefoil knot. G Li additionally contributed to the realization of non-Hermitian NKM in topoelectrical circuits. All authors discussed the theoretical and computational results and contributed to the writing of the manuscript.

\section*{Competing Interests}
The authors have no competing interests.
XZ is supported by the National Natural Science Foundation of China (Grant No. 11874431), the National Key R\&D Program of China (Grant No. 2018YFA0306800), and the Guangdong Science and Technology Innovation Youth Talent Program (Grant No. 2016TQ03X688). RT is supported by the excellence initiative DFG-EXC 2147/1 ct.qmat and the special research unit DFG-SFB 1170 tocotronics (project B04). CH Lee is supported by the Singapore Ministry of Education (MOE) Tier 1 start-up grant (R-144-000-435-133). GL is partially funded by Yat-Sen School, Sun Yat-Sen University.  TT is supported by the NSS scholarship by the Agency for Science, Technology and Research (A*STAR), Singapore.

%%%%%%%%%%%%%%%%%%%%%%%%%%%%%%%%%%%%%%%%%%%%%%%
%%%%%%%%%%%%%%%%%%%%%%%%%%%%%%%%%%%%%%%%%%%%%%%
%%%%%%%%%%%%%%%%%%%%%%%%%%%%%%%%%%%%%%%%%%%%%%%
%%%%%%%%%%%%%%%%%%%%%%%%%%%%%%%%%%%%%%%%%%%%%%%
\clearpage
\onecolumngrid
\appendix 
\section*{Supplementary Information}
\renewcommand{\figurename}{Supplemental Figure}
\renewcommand{\thefigure}{S\arabic{figure}}

\setcounter{figure}{0}
\author{Ching Hua Lee$^{1}$}
\email{phylch@nus.edu.sg}
\author{Guangjie Li$^{2,3}$}
\author{Yuhan Liu$^{2,4}$}
\author{Tommy Tai$^{5}$}
\author{Ronny Thomale$^6$}
\email{rthomale@physik.uni-wuerzburg.de}
\author{Xiao Zhang$^{2}$}
\email{zhangxiao@mail.sysu.edu.cn}
\affiliation{$^1$Department of Physics, National University of Singapore, Singapore 117542.}
\affiliation{$^2$School of Physics, Sun Yat-sen University, Guangzhou 510275, China.}
\affiliation{$^3$Department of Physics and Astronomy, Purdue University, IN 47907-2036, USA.}
\affiliation{$^4$Department of Physics, the University of Chicago, Chicago, Illinois 60637, USA.}
\affiliation{$^5$Cavendish Laboratory, University of Cambridge, JJ Thomson Ave, Cambridge CB3 0HE, United Kingdom}
\affiliation{$^6$Institut f\"{u}r Theoretische Physik und Astrophysik,
  Universit\"{a}t W\"{u}rzburg, Am Hubland Campus S\"{u}d,
  W\"{u}rzburg 97074, Germany}
\section{SUPPLEMENTARY NOTE 1}
In this section, we provide a more detailed analysis of the complex spectral properties i.e. imaginary gap and vorticity of the two models most featured in the main text, namely the Hopf-link and Trefoil nodal knot metals (NKMs).

\subsection{Hopf-Link NKM}
We analyze the tidal region of the simplest nodal knot metal (NKM) - the non-Hermitian Hopf-link. In Figs.~\ref{fig:XX2}a,b of the main text, the analytically derived topological tidal region from Eq.~\ref{tptm} of the main text Fig.~\ref{fig:XX2}(a) agrees exactly with the trenches (band intersections) in the imaginary band structure plot Fig.\ref{fig:XX2}(b), which define a tidal island. Note that other auxiliary peaks within the island, which possibly connect with other bands, play no role in topology. 

As a further elaboration of the marine landscape analogy, consider the periodic boundary conditions (PBC) scenario governed by Bloch states represented by the $\beta=\log|z|=0$ sea level, i.e. real $k_\perp=-i\log z$. As the boundary couplings are gradually switched off, a spectral flow ensues~\cite{PhysRevB.99.201103}, corresponding to a shift in the tidal (sea) level. The spectral flow stops at the trenches, which demarcate the ``actual tidal'' boundary of the topological region.

This spectral flow is laid out in detail in the figures that follow, Figs.~\ref{Hopfe1},\ref{Hopfe2},\ref{Hopfe3}, which feature the spectra at various representative surface momenta for each of the three possible surface terminations, $\hat{e}_1,\hat{e}_2,\hat{e}_3$ respectively. Since $h_z(\bold k)=0$ for all $\bold k$, topological boundary modes, if any, reside at the origin. Note that, due to limitations of numerical convergence, the numerical OBC (in black) spectra (under open boundary conditions) sometimes do not converge onto lines, even though they rigorously should in the OBC limit of exactly zero boundary hoppings. This illustrates that, under the skin effect, even infinitesimally small noise can significantly affect the spectrum.

\newpage
\begin{figure}[H]
\centering
\includegraphics[width=.9\linewidth]{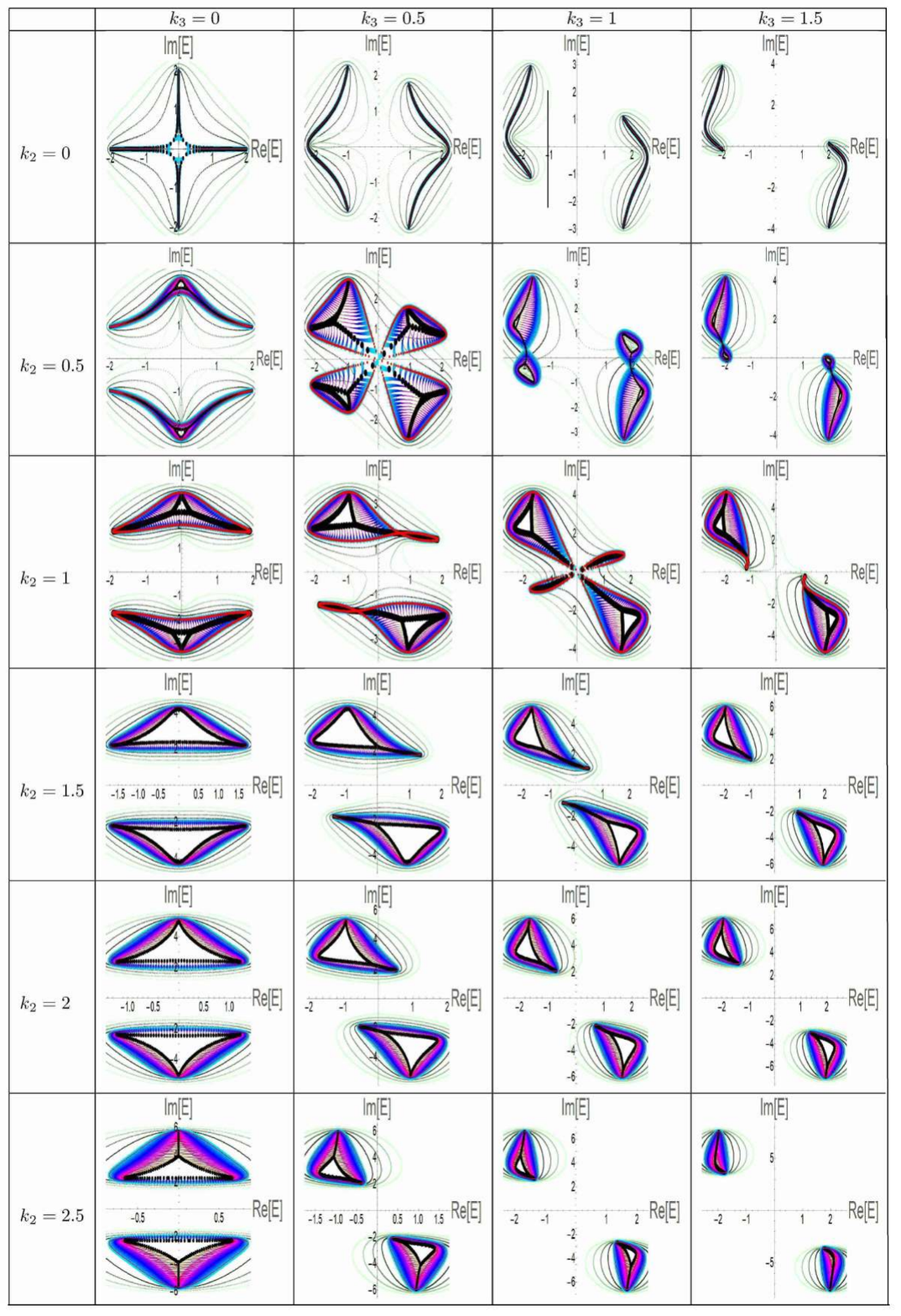}
\caption{Complex spectra of the non-Hermitian Hopf nodal knot metal (NKM) with $\hat e_1$ surface termination. Blue-magenta curves denote the spectral flow between the spectra under periodic boundary conditions (red) and open boundary conditions (black). Background contours denote level curves of constant $\log|z|$ where $z=e^{ik_1}$. }
\label{Hopfe1}
\end{figure}

\begin{figure}[H]
\centering
\includegraphics[width=.9\linewidth]{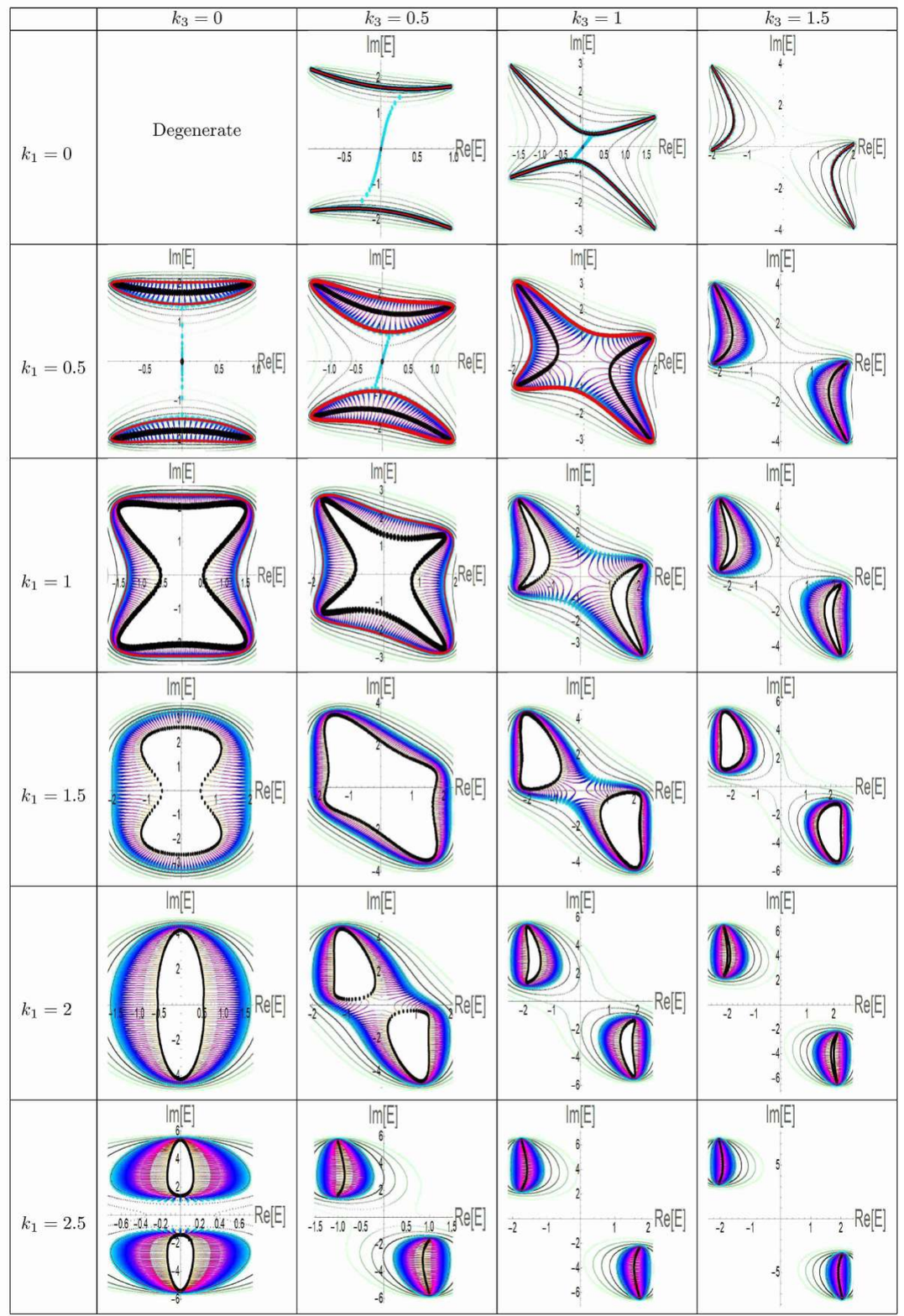}
\caption{Complex spectra of the non-Hermitian Hopf nodal knot metal (NKM) with $\hat e_2$ surface termination. Blue-magenta curves denote the spectral flow between the spectra under periodic boundary conditions (red) and open boundary conditions (black). Background contours denote level curves of constant $\log|z|$ where $z=e^{ik_2}$.}
\label{Hopfe2}
\end{figure}

\begin{figure}[H]
\centering
\includegraphics[width=.9\linewidth]{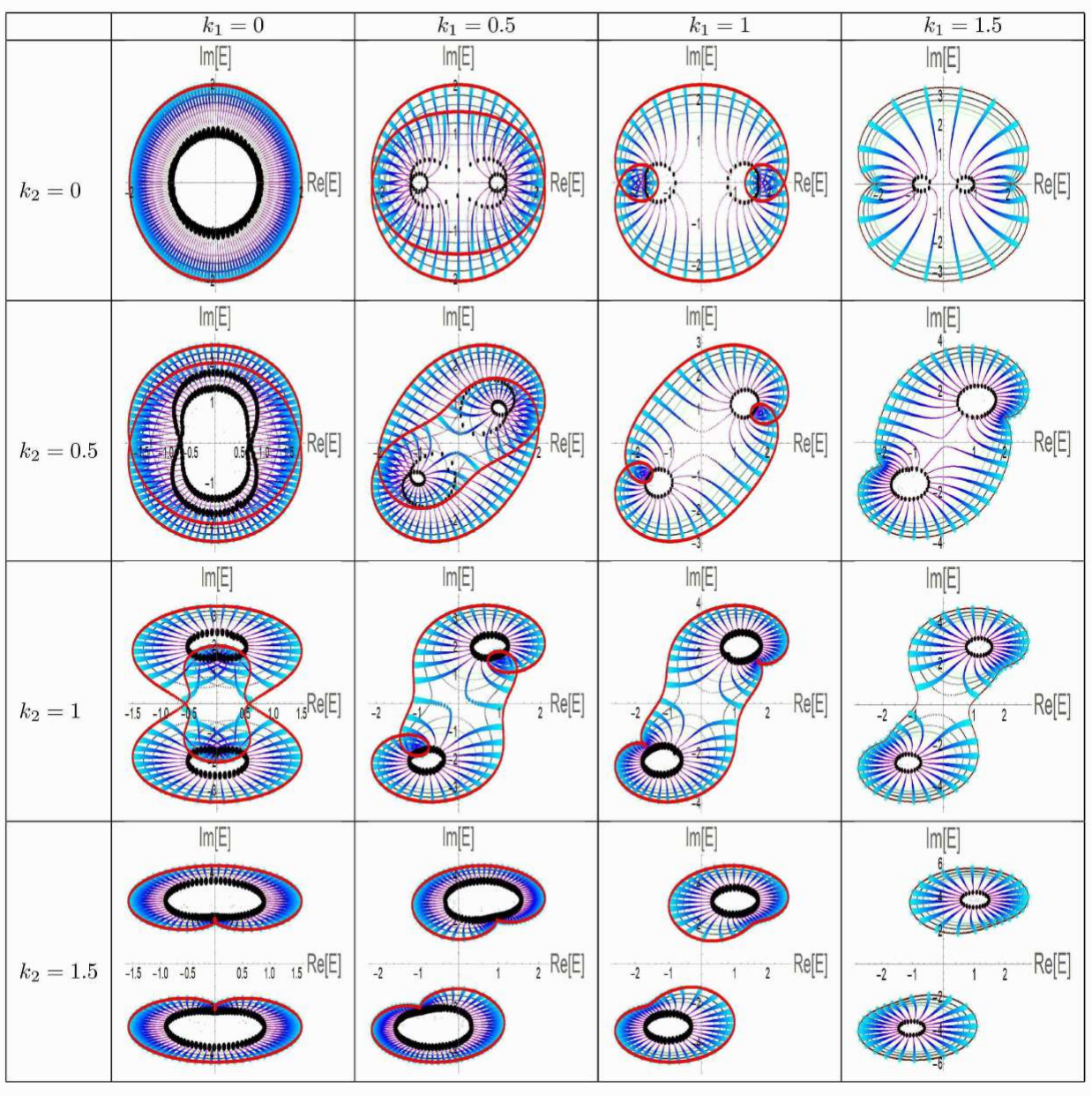}
\caption{Complex spectra of the non-Hermitian Hopf nodal knot metal (NKM) with $\hat e_3$ surface termination. Blue-magenta curves denote the spectral flow between the spectra under periodic boundary conditions (red) and open boundary conditions (black). Background contours denote level curves of constant $\log|z|$ where $z=e^{ik_3}$. Black circles represent the skin boundary states obtained at maximal numerical convergence; note the lack of well-defined curves of skin states.}
\label{Hopfe3}
\end{figure}

\newpage
\subsection{Trefoil NKM}
Here we provide further details of the complex analytic properties of our non-Hermitian Trefoil NKM model. 

Compared to the Hopf-link, the Trefoil NKM has a much richer complex gap band structure, with 8 solutions (bands). As explained and shown in Fig.~\ref{fig:XX2}(c-d) in the main text, only the intersection between the 4th and the 5th bands are of topological significance. Even so, there exist subtleties on the types of intersections that are actually significant. As explained in the main text, and present for the full surface Brillouin Zones (BZs) below, topological boundaries correspond only to the boundaries between the light and dark colored region in the plots of Fig.~\ref{TrefoilBandStructure} below. The other intersections, i.e. between light red and blue regions, also correspond to trenches, but not the tidal trenches of topological significance. 

Also presented in the following figures (Figs.~\ref{Trefoile1},\ref{Trefoile2},\ref{Trefoile3}) are the spectral flow plots, which are somewhat more intricate than those of the non-Hermitian Hopf model. Since $h_z(\bold k)=i\gamma$ with $\gamma$ set to unity, topological boundary modes, if any, occur at $\pm i$. Like in the Hopf-link case, limitations of numerical tolerance give rise to infinitesimally small noise in the boundary hoppings, which cause some OBC spectra (black) not to converge fully onto curves or straight lines.

\begin{figure}[H]
\centering
\includegraphics[width=.9\linewidth]{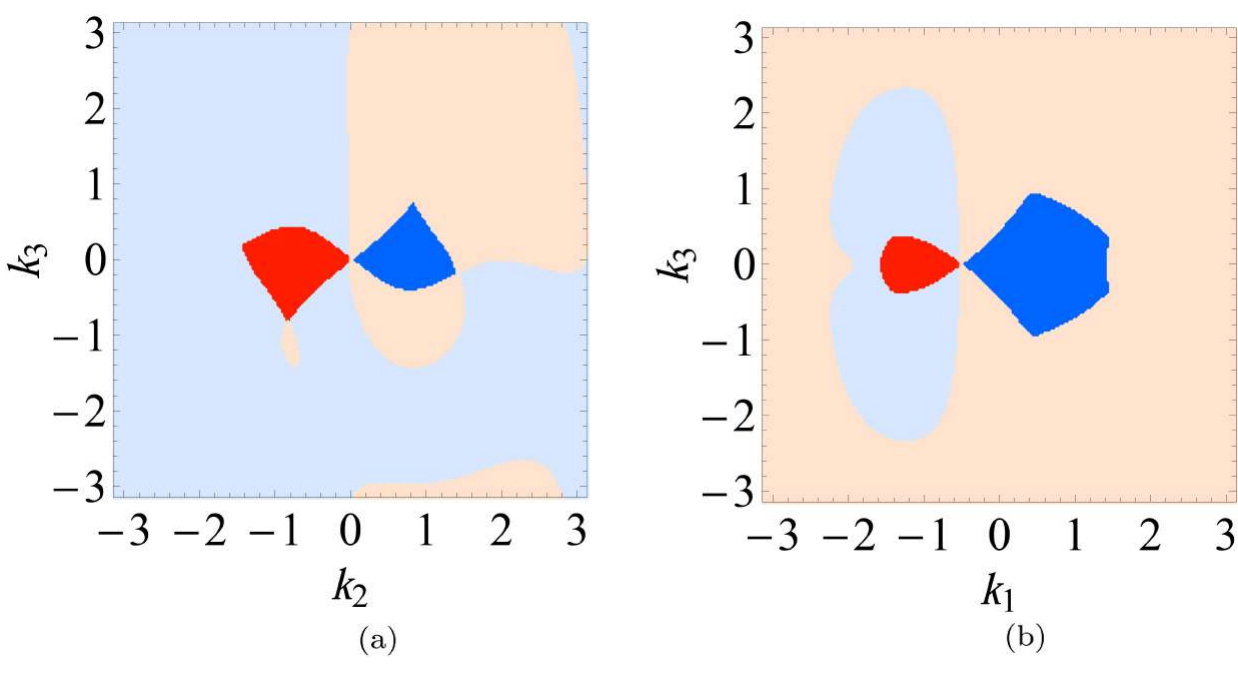}
\caption{Plots of the 4-th and 5-th $\log |z|$ (where $z=e^{ik_j}$, j=1,2,3) bands like in Fig.~\ref{fig:XX2}b of the main text, for the non-Hermitian Trefoil knot nodal knot metal with a) $\hat e_1$ and b) $\hat e_2$ surface terminations.}
\label{TrefoilBandStructure}
\end{figure}

\newpage
\begin{figure}[H]
\centering
\includegraphics[width=.9\linewidth]{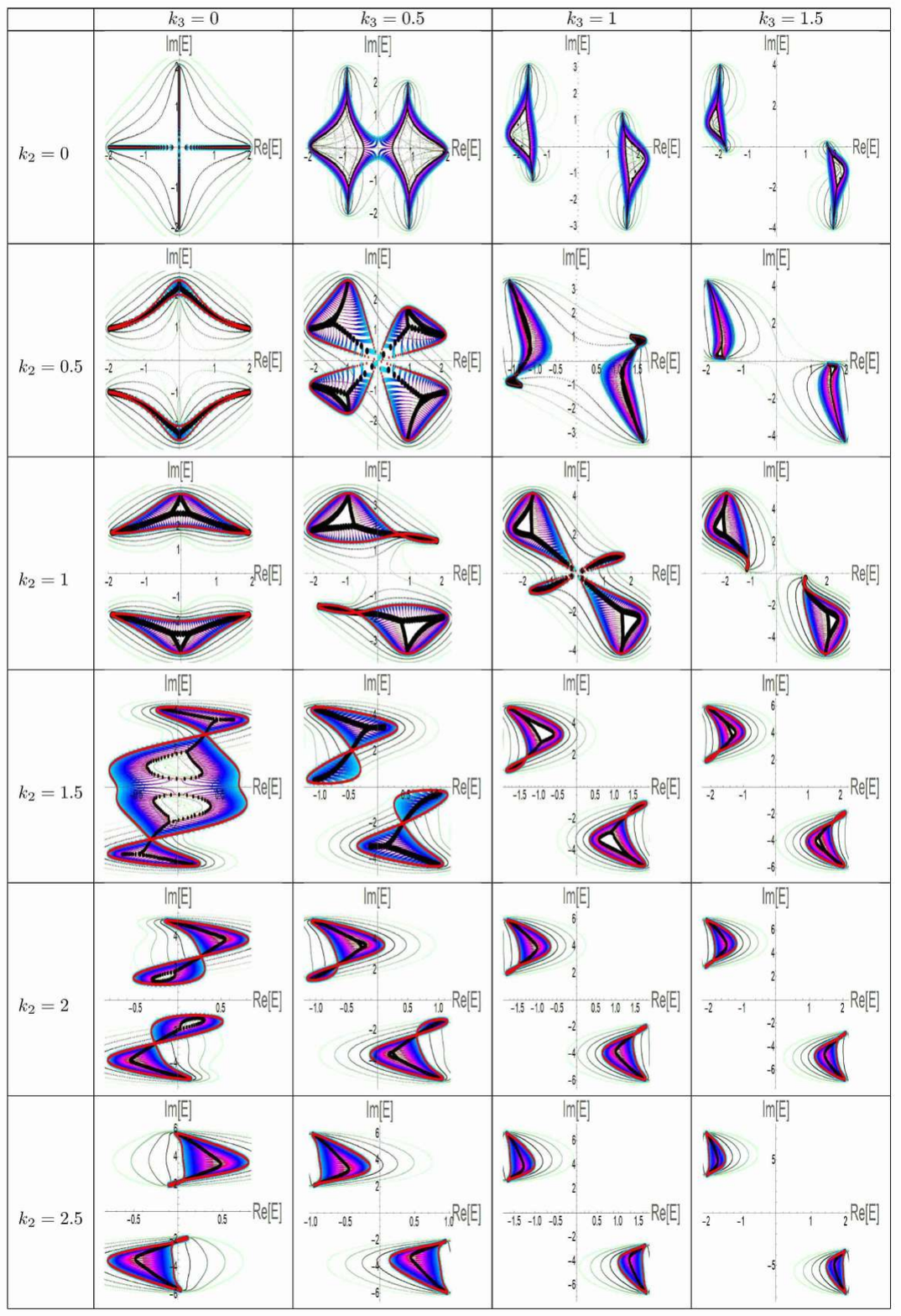}
\caption{Complex spectra of the non-Hermitian Trefoil nodal knot metal (NKM) with $\hat e_1$ surface termination. Blue-magenta curves denote the spectral flow between the spectra under periodic boundary conditions (red) and open boundary conditions (black). Background contours denote level curves of constant $\log|z|$ where $z=e^{ik_1}$.}
\label{Trefoile1}
\end{figure}

\begin{figure}[H]
\centering
\includegraphics[width=.9\linewidth]{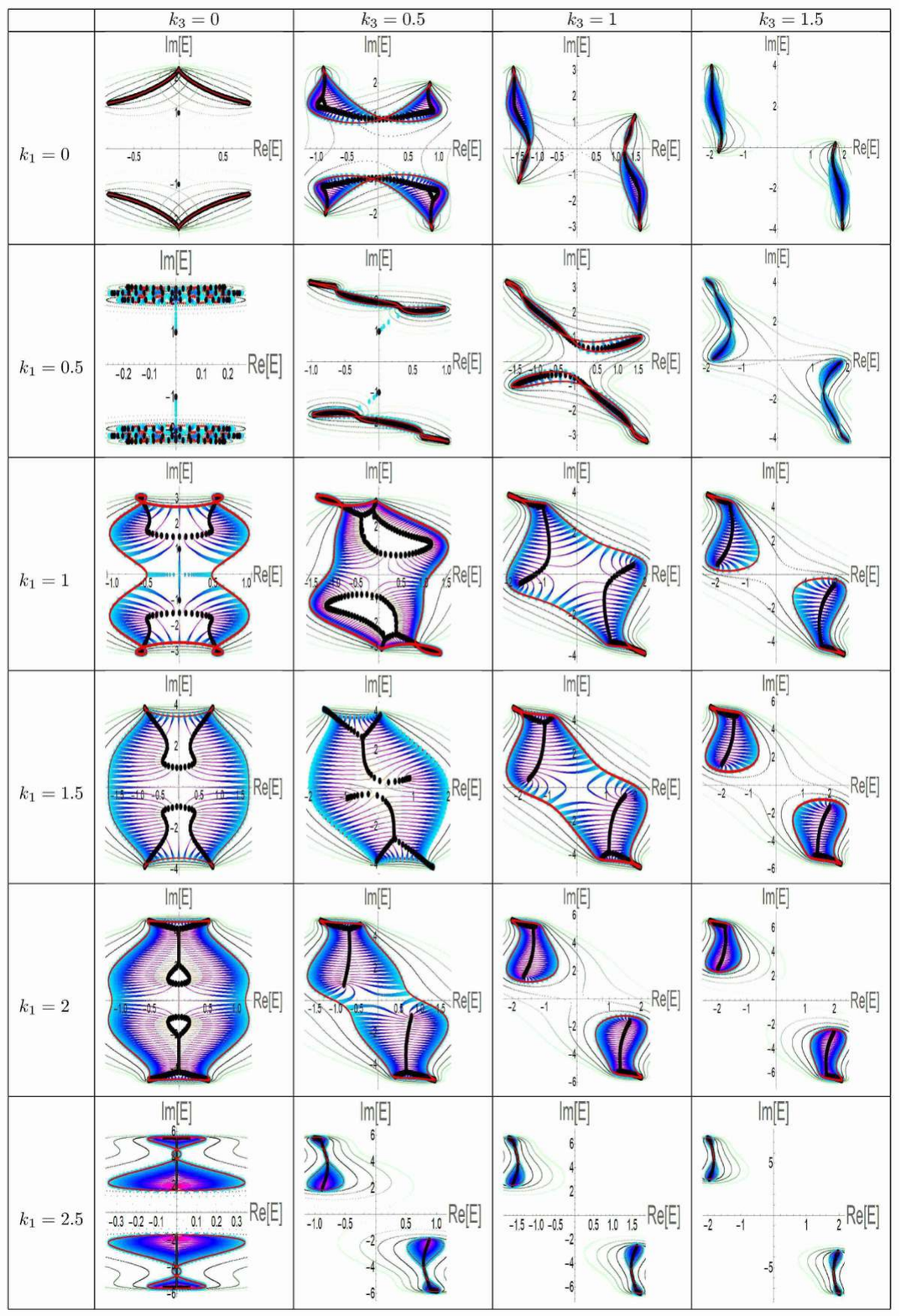}
\caption{Complex spectra of the non-Hermitian Trefoil nodal knot metal (NKM) with $\hat e_2$ surface termination. Blue-magenta curves denote the spectral flow between the spectra under periodic boundary conditions (red) and open boundary conditions (black). Background contours denote level curves of constant $\log|z|$ where $z=e^{ik_2}$.}
\label{Trefoile2}
\end{figure}

\begin{figure}[H]
\centering
\includegraphics[width=.9\linewidth]{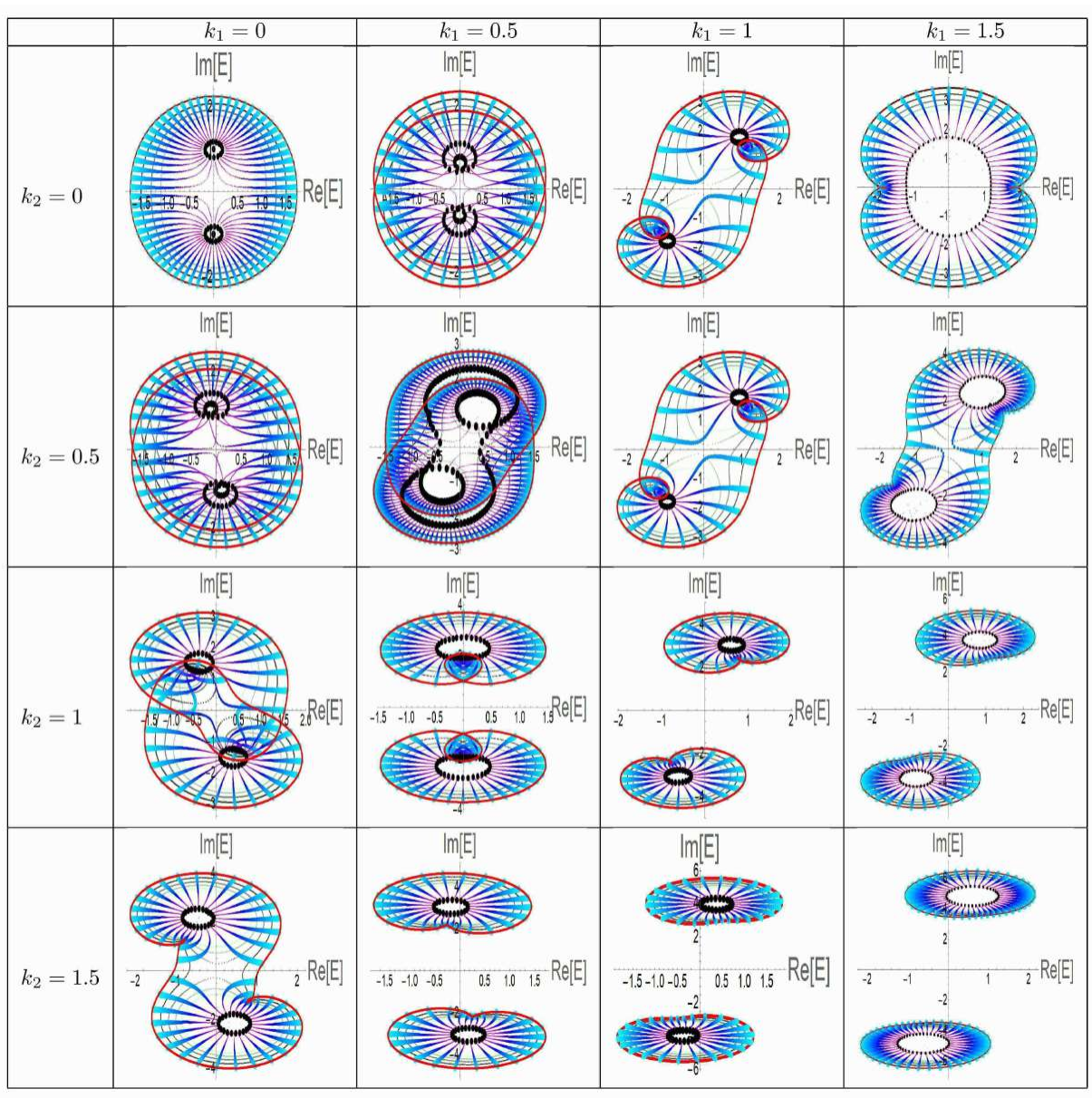}
\caption{Complex spectra of the non-Hermitian Trefoil nodal knot metal (NKM) with $\hat e_3$ surface termination. Blue-magenta curves denote the spectral flow between the spectra under periodic boundary conditions (red) and open boundary conditions (black). Background contours denote level curves of constant $\log|z|$ where $z=e^{ik_3}$. Like for the Hopf case, black loops represent the skin boundary states obtained at maximal numerical convergence.  }
\label{Trefoile3}
\end{figure}

\clearpage
\onecolumngrid
\section*{References}
\bibliography{references}

\end{document}